\documentclass[showpacs,amsmath,amssymb,twocolumn,superscriptaddress,aps,prb]{revtex4}
\usepackage{graphicx}
\usepackage{bbm}

\begin{document}

\title{Extended Ginzburg-Landau formalism: systematic expansion in small deviation from the critical temperature}

\author{A. V. Vagov}
\affiliation{Institut f\"{u}r Theoretische Physik III, Bayreuth
Universit\"{a}t, Bayreuth 95440, Germany}
\author{A. A. Shanenko}
\affiliation{Departement Fysica, Universiteit Antwerpen,
Groenenborgerlaan 171, B-2020 Antwerpen, Belgium}
\author{M. V. Milo\v{s}evi\'{c}}
\affiliation{Departement Fysica, Universiteit Antwerpen,
Groenenborgerlaan 171, B-2020 Antwerpen, Belgium}
\author{V. M. Axt}
\affiliation{Institut f\"{u}r Theoretische Physik III, Bayreuth
Universit\"{a}t, Bayreuth 95440, Germany}
\author{F. M. Peeters}
\affiliation{Departement Fysica, Universiteit Antwerpen,
Groenenborgerlaan 171, B-2020 Antwerpen, Belgium}

\date{\today}

\begin{abstract}
Based on the Gor'kov formalism for a clean $s$-wave superconductor, we develop an extended version of the single-band Ginzburg-Landau (GL) theory by means of a systematic expansion in the deviation from the critical temperature $T_c$, i.e., $\tau=1-T/T_c$. We calculate different contributions to the order parameter and the magnetic field: the leading contributions ($\propto \tau^{1/2}$ in the order parameter and $\propto \tau$ in the magnetic field) are controlled by the standard Ginzburg-Landau (GL) theory, while the next-to-leading terms ($\propto \tau^{3/2}$ in the gap and $\propto \tau^2$ in the magnetic field) constitute the extended GL (EGL) approach. We derive the free-energy functional for the extended formalism and the corresponding expression for the current density. To illustrate the usefulness of our formalism, we calculate, in a semi-analytical form, the temperature-dependent correction to the GL parameter at which the surface energy becomes zero, and analytically, the temperature dependence of the thermodynamic critical field. We demonstrate that the EGL formalism is not just a mathematical extension to the theory - variations of both the gap and the thermodynamic critical field with temperature calculated within the EGL theory are found in very good agreement with the full BCS results down to low temperatures, which dramatically improves the applicability of the formalism compared to its standard predecessor.
\end{abstract}

\maketitle

%%%%%%%%%%%%%%%%%%%%%%%%%%%%%%%%%%%%%%%%%%%%%%%%%%%%%%%%%%%%%%%%%%%%%%%%%%%%%%%%%%%%
\section{Introduction}
\label{sec:introduction}
%%%%%%%%%%%%%%%%%%%%%%%%%%%%%%%%%%%%%%%%%%%%%%%%%%%%%%%%%%%%%%%%%%%%%%%%%%%%%%%%%%%%

In 1950 Ginzburg and Landau proposed a phenomenological theory of superconductivity (the GL theory) based on a specific form of the free energy functional constructed in the vicinity of the critical temperature from Landau's theory of second-order
transitions.~\cite{GL} Minimization of this functional generates the system of two GL equations that give the spatial distribution of the superconducting order parameter (the condensate wave function) and of the magnetic field in a superconductor. Over the years, the GL approach has been enormously successful in describing various properties of superconductors (see e.g. Ref.~\onlinecite{degen}), and has been extensively used particularly in the last decade in the domain of mesoscopic superconductivity, see, e.g., Ref.~\onlinecite{mesoGL}.

In 1957 Bardeen, Cooper and Schrieffer (BCS) developed the well-known microscopic theory of the conventional superconductivity~\cite{BCS} and then, two years later, Gor'kov showed that the GL equations can be obtained from the BCS formalism.~\cite{gor} However, in spite of the availability of the microscopic theory, the GL approach remains still an optimal choice for many practical calculations when the spatial distribution of the pair condensate and, thus, of the magnetic field are nontrivial, e.g., for multiple-vortex configurations. The simple differential structure of the local GL equations in comparison with the microscopic theory offers clear advantages, including the possibility of analytical derivations in many important cases.

A desire to develop an extension to the GL theory, with the idea of improving the formalism while retaining at least some advantages of its original formulation, stimulated significant efforts by many theorists. Several GL-like theories of different complexity were proposed. In the earliest developments,~\cite{tewordt,werthammer} the so-called ``local superconductor'' formalism was attempted, being a complicated synthesis of the BCS and GL approaches. The GL theory with nonlocal corrections (i.e., including higher powers of the gradients of the order parameter) was used in studies of the anisotropy of the upper critical field (see Ref.~\onlinecite{takanaka}) and the vortex structure (see Ref.~\onlinecite{ichioka}) in $d$-wave superconductors. Recently, various extensions to the GL theory were introduced in the context of studying the Fulde-Ferrell-Larkin-Ovchinnikov state.~\cite{buzdin,adachi,mineev} In all the above examples, extending the GL theory was based on the expansion of the  self-consistent gap equation by including higher powers of the order parameter and its spatial gradients phenomenologically.~\cite{note1} However, accounting for such higher-power terms is not as straightforward as it may seem. The fundamental problem here is to select properly all relevant contributions of the same order of magnitude (accuracy). This (serious) issue does not arise in the derivation of the original GL theory for a single-band superconductor, where only the first non-linear term and the second-order (leading) gradient of the condensate wave function are included. However, as recently shown in Ref.~\onlinecite{kogan}, a similar selection performed for the GL theory of a two-band superconductor leads to the appearance of incomplete higher-order contributions. Such incomplete terms may cause misleading conclusions and should be avoided.

To tackle this problem, one needs to work with a single small parameter in the expansion. In the present case, such a small parameter is the proximity to the critical temperature, i.e., $\tau = 1-T/T_c$. Indeed, the standard GL approach can be seen as the lowest-order theory in the $\tau$-expansion of the self-consistent gap equation, see, e.g., Refs.~\onlinecite{kogan} and \onlinecite{shanenko}. However, next orders in $\tau$ are also of great importance, for example, to capture the physics of different healing lengths of different condensates in multi-band superconductors.~\cite{shanenko,lucia} In the present paper, we show that next orders in $\tau$ are also important in the single-band case, surprisingly improving the GL theory. We obtain a systematic series expansion of the self-consistent gap equation for a single-band, $s$-wave, clean superconductor, using $\tau$ as the governing small parameter. In the derivation we employ a technique in the spirit of the asymptotic expansion methods used extensively in the applied mathematics.~\cite{babich} Similarly to the asymptotic analysis in other models, we obtain a hierarchical system of the so-called transport equations which need to be solved recursively starting from the lowest order. Using this method, corrections to the standard GL theory can in principle be calculated with an arbitrary accuracy. However, unlike asymptotic expansions for linear models, here the complexity of the higher-order equations increases rapidly and their solution cannot be obtained in the general form. Based on the Gor'kov Green function formalism, we derive and investigate the first three orders of the $\tau$-expansion of the gap equation, i.e., $\tau^{n/2}$ with $n=1,2,3$. To the order $\tau^{1/2}$, we find the equation for the critical temperature. Collecting the terms proportional to $\tau^{3/2}$, we obtain the standard GL theory giving the lowest-order (in $\tau$) contributions to the superconducting condensate, i.e., $\propto \tau^{1/2}$, and to the magnetic field, i.e., $\propto \tau$. Then, by matching the terms of the order $\tau^{5/2}$, we derive equations for the next-to-leading corrections to the superconducting order parameter and magnetic field, $\propto \tau^{3/2}$ and $\tau^2$, respectively. The equations controlling the order parameter and the magnetic field up to the next-to-leading order in $\tau$ constitute the extended GL formalism (EGL). To illustrate the power of our extension to the GL theory, we investigate the energy associated with a surface between the superconducting (S) and normal (N) phases. In particular, we calculate the temperature dependence of $\kappa^*$, the value of the GL parameter at which the surface energy becomes zero. It is important to stress here that contrary to other available extensions of the GL theory discussed above, our formalism is not much more complicated than the standard GL theory. As is shown in the calculation of the S/N surface energy, plenty of important information can be obtained from the EGL formalism analytically.

Note that as is known, the GL theory is heavily used also in the study of thermal fluctuations. Here we do not address this issue but deal with an extension to the GL formalism in the mean-field level. Our aim is to expand the validity domain of the GL theory down to lower temperatures, which will be useful for the problems with a strongly nonuniform distribution of the pair condensate, e.g., for multiple vortex solution in the presence of stripes and clusters of vortices. In this case any fully microscopic approach will be an very complicated and rather time consuming task.

The paper is organized as follows. In Sec.~\ref{sec:tau-expansion-zero} we introduce a general approach to construct the asymptotic expansion in $\tau$ for the self-consistent gap equation. In order to illustrate the main ideas behind the method, a simpler case of zero magnetic field is considered here. Sec.~\ref{sec:tau-expansion-nonzero} presents the generalization to a nonzero magnetic field. Such a generalization requires the normal-metal Green function beyond the traditional Peierls (phase) approximation, and the corresponding expression is presented and discussed. The series expansion of the free-energy functional up to the next-to-leading order in $\tau$ is given in Sec.~\ref{sec:free-energy}. The complete set of equations for the order parameter and the magnetic field in the EGL approach is derived by finding the stationary point of the functional in Sec.~\ref{sec:complete-equations}. In Sec.~\ref{sec:GL_domain} we estimate the accuracy of the EGL formalism by comparing its results for the uniform order parameter and the critical magnetic field with those of the standard GL approach and the BCS theory. Here we demonstrate that the temperature in which the GL theory is valid, is dramatically increased due to the extension. In Sec.~\ref{sec:surface-energy} we investigate the S/N surface energy and calculate the temperature-dependent correction to $\kappa^*$. Finally, Sec.~\ref{sec:conclusions} presents the summary of our results and our conclusions.

The main text of the paper contains only the key formulas, the general ideas and the main steps of the derivation. Readers interested in details are referred to Appendices. In particular, Appendix~\ref{sec:appendix_A} shows how to calculate the coefficients appearing in the $\tau$-expansion of the gap equation in the absence of a magnetic field. Appendix~\ref{sec:appendix_B} presents details of our calculations for the normal-metal Green function beyond the Peierls phase approximation. Appendix~\ref{sec:appendix_C} generalizes the calculations given in Appendix~\ref{sec:appendix_A} to the case of a nonzero magnetic field.

%%%%%%%%%%%%%%%%%%%%%%%%%%%%%%%%%%%%%%%%%%%%%%%%%%%%%%%%%%%%%%%%%%%%%%%%%%%%%%%%
\section{Series expansion in $\tau$ of the gap equation at zero magnetic field}
\label{sec:tau-expansion-zero}
%%%%%%%%%%%%%%%%%%%%%%%%%%%%%%%%%%%%%%%%%%%%%%%%%%%%%%%%%%%%%%%%%%%%%%%%%%%%%%%%

As is known since the classical work by Gor'kov,~\cite{gor,degen,fett} the GL equations can be derived from the microscopic BCS theory in the most elegant way via the Green function formalism. For the sake of clarity of presentation of our main ideas, the case of zero magnetic field is considered first, while the generalization to a nonzero magnetic field is given in the next section. The goal of our work is to construct the extended GL formalism through the expansion of the Gor'kov equations in $\tau = 1- T/T_c$, with $T_c$ the critical temperature ($T < T_c$). We start by writing the Gor'kov equations as the Dyson equation for the Green function ${\check {\cal G}}_{\omega}$ in the Gor'kov-Nambu $2\time2-$matrix representation, which reads as (see, e.g., Ref.~\onlinecite{fett,zagoskin,kopnin})
\begin{equation}
{\check {\cal G}}_{\omega}={\check {\cal G}}^{(0)}_{\omega} +
{\check {\cal G}}^{(0)}_{\omega}\;{\check \Delta}\;{\check {\cal
G}}_{\omega}, \label{Geq}
\end{equation}
with
\begin{equation}
{\check {\cal G}}_{\omega}= \left(
\begin{array}{cc}
{\cal G}_{\omega} & {\cal F}_{\omega}\\
{\widetilde {\cal F}}_{\omega} & {\widetilde {\cal G}}_{\omega}
\end{array}
\right), \quad {\check {\cal G}}^{(0)}_{\omega}= \left(
\begin{array}{cc}
{\cal G}^{(0)}_{\omega} & 0\\
0 & {\widetilde {\cal G}}^{(0)}_{\omega}
\end{array}
\right), \label{Gfunc}
\end{equation}
where $\hbar\omega= \pi T(2n+1)$ is the fermionic Matsubara frequency ($n$ is an integer and $k_B$ is set to unity) and the $2\times2$ matrix gap operator ${\check \Delta}$ in Eq.~(\ref{Geq}) is defined by
\begin{equation}
{\check \Delta}= \left(
\begin{array}{cc}
0 & {\hat \Delta}\\
{\hat \Delta}^{\ast} & 0
\end{array}
\right), \quad \langle {\bf r}|{\hat \Delta}|{\bf r}'\rangle=
\delta({\bf r}-{\bf r}') \Delta({\bf r}'), \label{gapcheck}
\end{equation}
where the superconducting order parameter $\Delta({\bf r})$ obeys the self-consistent gap equation, i.e.,
\begin{align}
\Delta({\bf r}) = - gT\sum\limits_\omega {\cal F}_\omega({\bf r},{\bf r}),
\label{eq:gap}
\end{align}
with $g$ the (Gor'kov) coupling constant. As usual, the sum in the r.h.s. of
Eq.~(\ref{eq:gap}) is assumed to be properly restricted to avoid the ultraviolet divergence. Equations (\ref{Geq}) and (\ref{Gfunc}) further give
\begin{subequations}
\label{eq:Gorkov_equations}
\begin{align}
&{\cal F}_{\omega}({\bf r},{\bf r}')=\int\!\!{\rm d}^3y
\;{\cal G}^{(0)}_{\omega}({\bf r},{\bf y})\,\Delta({\bf y})\,
{\widetilde {\cal G}}_{\omega}({\bf y},{\bf r}'),\\
&{\widetilde {\cal G}}_{\omega}({\bf r},{\bf r}')=
{\widetilde {\cal G}}^{(0)}_{\omega}({\bf r},{\bf r}') \nonumber\\
&\quad\quad\quad\quad\quad+ \int\!\!{\rm d}^3y\;{\widetilde {\cal
G}}^{(0)}_{\omega}({\bf r},{\bf y})\;{\hat \Delta}^{\ast}({\bf y})\,
{\cal F}_{\omega}({\bf y},{\bf r}').
\end{align}
\end{subequations}
For the normal-state temperature Green function ${\cal G}_{\omega}^{(0)}({\bf r},{\bf y})$ we have (at zero magnetic field)
\begin{align}
{\cal G}_{\omega}^{(0)}({\bf r}, {\bf y}) = \int\!\!\frac{{\rm
d}^3k}{(2 \pi)^3}\;\frac{e^{i {\bf k} ({\bf r}-{\bf
y})}}{i\hbar\omega - \xi_k}, \label{eq:green_0}
\end{align}
with the single-particle energy $\xi_k = \hbar^2 k^2/2m - \mu$ measured from the chemical potential $\mu$, and $\widetilde {\cal G}_{\omega}({\bf r},{\bf y}) =-{\cal G}_{-\omega}({\bf y},{\bf r})$.

The Gor'kov equations (\ref{eq:Gorkov_equations}) supplemented by Eq.~(\ref{eq:green_0}) make it possible to express the anomalous (Gor'kov) Green function ${\cal F}_{\omega}({\bf r},{\bf r}')$ in terms of $\Delta({\bf r})$ and
${\cal G}_{\omega}^{(0)}({\bf r}, {\bf y})$. Then, inserting this expression into Eq.~(\ref{eq:gap}), one obtains the self-consistent gap equation. Solution to the gap equation can be represented in the form of a perturbation series over powers of $\Delta({\bf r})$~(which is small in the vicinity of the critical temperature
$T_c$):
\begin{align}
\Delta ({\bf r}) & = \int\!\!{\rm d}^3y\, K_a({\bf r},{\bf y})
\Delta({\bf y}) +\int\!\!\prod_{j=1}^3{\rm d}^3y_j \;K_b({\bf r},
\{{\bf y}\}_3) \notag \\& \times \Delta({\bf y}_1)\Delta^{\ast}({\bf
y}_2)\Delta({\bf y}_3) + \int\!\!\prod_{j=1}^5 {\rm d}^3y_j \;
K_c({\bf r},\{{\bf y}\}_5) \notag \\
& \times \Delta({\bf y}_1)\Delta^{\ast}({\bf y}_2)\Delta({\bf
y}_3)\Delta^{\ast}({\bf y}_4)\Delta({\bf y}_5) + \dots,
\label{eq:gap_perturbation}
\end{align}
where $\{{\bf y}\}_n = \{{\bf y}_1,\ldots,{\bf y}_n\}$ and the integral kernels are given by
\begin{align}
& K_a({\bf r},{\bf y} ) = - gT\,\sum\limits_{\omega}\,{\cal G}^{
(0)}_{\omega}({\bf r}, {\bf y}){\widetilde {\cal G}}^{(0)}_{
\omega}({\bf y},{\bf r}),\notag\\
& K_b({\bf r},\{{\bf y}\}_3) = -gT\,\sum\limits_{\omega}\,{\cal
G}^{(0)}_{\omega}({\bf r},{\bf y}_1){\widetilde {\cal G}}^{(0)}_{
\omega}({\bf y}_1, {\bf y}_2)  \notag \\
&\quad\quad\quad\quad\quad\quad\quad\quad\quad\quad\times {\cal
G}^{(0)}_{\omega}({\bf y}_2,{\bf y}_3){\widetilde {\cal
G}}^{(0)}_{\omega} ({\bf y}_3,{\bf r}),
\label{eq:kernels}\\
&  K_c ({\bf r},\{{\bf y}\}_5)=-gT\,\sum\limits_{\omega}{\cal
G}^{(0)}_{\omega}({\bf r},{\bf y}_1){\widetilde {\cal G}}^{(0)}_{
\omega}({\bf y}_1,{\bf y}_2)\notag \\
&\times {\cal G}^{(0)}_{\omega} ({\bf y}_2,{\bf y}_3){\widetilde
{\cal G}}^{(0)}_{\omega} ({\bf y}_3,{\bf y}_4){\cal
G}^{(0)}_{\omega} ({\bf y}_4,{\bf y}_5){\widetilde {\cal
G}}^{(0)}_{\omega}({\bf y}_5, {\bf r}). \notag
\end{align}
Equation~(\ref{eq:gap_perturbation}) can be truncated to a desired order, which yields a non-linear integral equation. The latter is further converted into a non-linear partial differential equation by using the gradient expansion
\begin{align}
\Delta({\bf y}_j)=\Delta ({\bf r} + {\bf z}_j ) = \sum_{n=0}^{\infty}\frac{1}{n!} \big({\bf z}_j \cdot {\boldsymbol \nabla}_{{\bf r}}\big)^n \Delta ({\bf r}),
\label{eq:gradient_expansion}
\end{align}
with ${\bf z}_j\cdot {\boldsymbol \nabla}_{{\bf r}}$ the scalar product. In particular, the GL equation is obtained when keeping only the first two terms, including $K_a$ and $K_b$, in Eq.~(\ref{eq:gap_perturbation}) and the second-order spatial derivatives in the gradient expansion (\ref{eq:gradient_expansion}).

While Eq.~(\ref{eq:gap_perturbation}) is a regular series expansion of the gap equation (\ref{eq:gap}), the partial differential equation mentioned above is not. The gradient expansion introduces a second small parameter together with the corresponding truncation approximation, and the relation between the order parameter and its gradients is not known {\it a priory}. As a result one cannot compare the accuracy of the relevant terms in the expansion and the truncation procedure becomes ill-defined. This problem does not appear in the derivation of the GL equation where one keeps only the lowest second order gradient term. In order to extend the GL formalism, one has to deal with a single small parameter in the system that can be used to compare the relevant contributions. This small parameter follows from the solution of the GL equation. When $T \rightarrow T_c$ the order parameter decays as $\Delta \propto \tau^{1/2} \rightarrow 0$. Also the solution reveals the scaling length $\xi \propto \tau^{-1/2}$, i.e., the GL coherence length, which determines the spatial variations of the order parameter in the vicinity of $T_c$ and dictates that ${\boldsymbol \nabla}\Delta \propto \tau$, or, using the short-hand notation, ${\boldsymbol \nabla} \propto \tau^{1/2}$. Thus, the parameter $\tau$ controls both relevant quantities and can be used to produce a single-small-parameter series expansion of the gap equation~(\ref{eq:gap}).

A systematic expansion of the gap equation in $\tau$ can be facilitated by introducing the scaling transformation for the order parameter, the coordinates and the spatial derivatives of the order parameter in the following form:
\begin{align}
\Delta=\tau^{1/2} \bar{\Delta}, \quad {\bf r}=\tau^{-1/2} \bar{{\bf r}}, \quad {\boldsymbol \nabla}_{\bf r}=\tau^{1/2} {\boldsymbol \nabla}_{\bar{\bf r}}.
\label{eq:scaling}
\end{align}
Note that in terms of the scaled coordinates, the typical spatial variation of the order parameter occurs on a scale that is independent of $\tau$ to the leading order. After the transformation given by Eq.~(\ref{eq:scaling}), the parameter $\tau$ enters the expansion in Eqs.~(\ref{eq:gap_perturbation}) and (\ref{eq:gradient_expansion}) as follows. In Eq.~(\ref{eq:gradient_expansion}), only coordinate ${\bf r}$ changes. The scaling of the difference ${\bf z}$ does not change the expressions as it is an integration variable, and thus the scaling will
not be applied to it. As $\Delta$ is now a function of $\bar{\bf r}$, each derivative ${\boldsymbol \nabla}$ in the expansion (\ref{eq:gradient_expansion}) adds a factor
$\tau^{1/2}$. Inserting Eq.~(\ref{eq:gradient_expansion}) with these factors into the expansion given by Eq.~(\ref{eq:gap_perturbation}) and taking into account the factor $\tau^{1/2}$ for the order parameter in Eq.~(\ref{eq:scaling}) we arrive at a simple mnemonic rule to count the minimal order of each term in the expansion of the gap equation: {\it each occurrence of $\Delta$ or ${\boldsymbol \nabla}$ in the formulas adds the factor $\tau^{1/2}$}. The final form of the expansion is obtained by calculating the relevant coefficients through the evaluation of the remaining integrals. As those coefficients depend on temperature, they can also represented as series in $\tau$. The formulated procedure allows one to calculate the $\tau$-expansion for the gap equation to arbitrary order. However, in practice, calculations of higher orders become more and more complicated. In this work we limit ourselves to the analysis of the self-consistent gap equation in the first three orders, i.e., up to the order $\tau^{5/2}$. Collecting terms of the order $\tau^{1/2}$, we obtain the equation for $T_c$. Working in the order $\tau^{3/2}$, we recover the standard GL theory producing the leading contribution to $\Delta$, i.e., $\propto \tau^{1/2}$. The order $\tau^{5/2}$ yields the equation that controls the next-to-leading contribution to $\Delta$, i.e., $\propto \tau^{3/2}$~(this is what we call the EGL formalism). Details of the selection of {\it all the necessary terms} in Eq.~(\ref{eq:gap_perturbation}) that contribute to one of the three orders mentioned above, are given in Appendix~\ref{sec:appendix_A}. The final result reads ($\bar{\boldsymbol \nabla} ={\boldsymbol \nabla}_{\bar{\bf r}}$)
\begin{align}
\frac{\tau^{1/2}}{g}&\bar \Delta  =  a_1\tau^{1/2} \bar{\Delta}  +
a_2\tau^{3/2}\bar{\bf \nabla}^2\bar{\Delta} + a_3\tau^{5/2} \bar{\bf
\nabla}^2(\bar{\bf \nabla}^2 \bar{\Delta})\notag \\
-&b_1 \tau^{3/2} |\bar{\Delta}|^2 \bar{\Delta} - b_2\tau^{5/2} \Big[
2\bar{\Delta}\;|\bar{\bf {\boldsymbol \nabla}} \bar{\Delta}|^2 + 3 \bar{\Delta}^\ast
(\bar{\bf {\boldsymbol \nabla}} \bar{\Delta})^2 \nonumber \\
+&\bar{\Delta}^2\; \bar{\bf \nabla}^2\bar{\Delta}^{\ast}
+4|\bar \Delta|^2 \bar{\bf \nabla}^2\bar{\Delta}\Big]
+ c_1 \tau^{5/2}  |\bar \Delta|^4 \bar \Delta ,
\label{eq:gap_equation_series}
\end{align}
where the coefficients $a_i$, $b_i$ and $c_i$ are obtained from the integrals with the kernels $K_a$, $K_b$ and $K_c$, respectively, and they are given by
\begin{align}
&a_1 = {\cal A}_T - a\Big(\tau +\frac{\tau^2}{2}+ {\cal O}(\tau^3)
\Big), \;\frac{{\cal A}_T}{N(0)}=\ln\Big(\frac{2e^\gamma\hbar
\omega_D}{\pi T_c}\Big),\nonumber\\
&b_1 = b\left(1 + 2\tau + {\cal O}(\tau^2)\right), \quad b=\,N(0)
\frac{7\zeta(3)}{8\pi^2T_c^2}, \nonumber\\
&c_1 = c \big(1+{\cal O}(\tau)\big), \quad c=N(0)\,\frac{93\zeta(5)}{128
\pi^4T_c^4}, \nonumber\\
&a_2 = {\cal K}\left(1 + 2\tau+{\cal O}(\tau^2)\right), \quad {\cal K} =
\frac{b}{6}\hbar^2 v_F^2, \nonumber\\
&a_3 = {\cal Q} \big(1+{\cal O}(\tau)\big),\quad {\cal Q}=\frac{c}{30}
\hbar^4 v_F^4, \nonumber \\
&b_2= {\cal L}\big(1 + {\cal O}(\tau)\big),\quad {\cal L} = \frac{c}{9}
\hbar^2 v_F^2,
\label{eq:coefficients}
\end{align}
where $a=-N(0)$ and $N(0) = mk_F/(2\pi^2 \hbar^2)$ is the DOS at the Fermi energy, with $v_F$ the Fermi velocity; $\omega_D$ denotes the Debye (cut-off) frequency, $\gamma=0.577$ is the Euler constant and $\zeta(\ldots)$ is the Riemann zeta-function. It is of importance to note that Eq.~(\ref{eq:gap_equation_series}) contains only half-integer powers of $\tau$. The reason for this is two-fold. First, due to the structure of Eq.~(\ref{eq:gap_perturbation}), there appear only odd integer powers of the order parameter. Second, the spherical symmetry dictates that the integrals with an odd number of ${\boldsymbol \nabla}$ operators are equal to zero.

The solution to the gap equation (\ref{eq:gap_equation_series}) must also be sought in the form of a series expansion in $\tau$. Based on  Eqs.~(\ref{eq:gap_equation_series}) and (\ref{eq:coefficients}), we are able to introduce
\begin{align}
\bar{\Delta}({\bf r}) = \bar{\Delta}_0({\bf r}) + \tau \bar{\Delta}_1
({\bf r}) + \dots.
\label{eq:solution_expansion}
\end{align}
Substituting this into Eq.~(\ref{eq:gap_equation_series}) and collecting terms of the same order we obtain a set of equations for each $\Delta_n$.

Collecting terms of the order $\tau^{1/2}$, we obtain
\begin{align}
\big(g^{-1} - {\cal A}_T\big)\bar{\Delta}_0=0.
\label{eq:tau_1_2}
\end{align}
The solution to this equation, i.e., $g{\cal A}_T = 1$ gives the ordinary BCS expression for the critical temperature, i.e., $T_c = (2e^{\gamma}/\pi)\hbar \omega_D\exp[-1/gN(0)]$.

In the order $\tau^{3/2}$ one recovers the standard GL equation for the leading contribution to the order parameter $\Delta_0$:
\begin{align}
a\bar{\Delta}_0 + b|\bar{\Delta}_0|^2 \bar{\Delta}_0 - {\cal K}
\bar{\nabla}^2\bar{\Delta}_0 = 0.
\label{eq:GLE}
\end{align}
Note that its standard form with the temperature dependent $a$-coefficient is obtained by multiplying all terms by the factor $\tau^{3/2}$ and returning to the unscaled quantities, i.e.,
\begin{align}
a \tau \Delta_0 + b|\Delta_0|^2 \Delta_0 - {\cal K} \nabla^2\Delta_0 = 0.
\nonumber
\end{align}

Finally, collecting all terms of the order $\tau^{5/2}$, we arrive at the equation for $\Delta_1$, i.e., the next-to-leading term in the order parameter:
\begin{align}
a\bar{\Delta}_1 + b\big(2|\bar{\Delta}_0|^2\bar{\Delta}_1+\bar{\Delta}_0^2
\bar{\Delta}^{\ast}_1\big)-{\cal K} \bar \nabla^2\bar\Delta_1= F,
\label{eq:GLE_corr}
\end{align}
where $F$ is given by
\begin{align}
F = &-\frac{a}{2} \bar{\Delta}_0 +2 {\cal K} \bar{\bf \nabla}^2\bar{\Delta}_0
+ {\cal Q} \bar{\bf \nabla}^2 \big(\bar{\bf \nabla}^2\Delta_0\big)
\nonumber \\
&-2b|\bar{\Delta}_0|^2\bar{\Delta}_0 - {\cal L}\Big[2\bar{\Delta}_0\;|\bar{\boldsymbol \nabla}\bar{\Delta}_0|^2+3\bar{\Delta}_0^\ast
\;\big(\bar{\boldsymbol \nabla}\bar{\Delta}_0\big)^2\nonumber\\
&+\bar{\Delta}_0^2\;\bar{\bf \nabla}^2\bar{\Delta}^{\ast}_0 +
4|\bar{\Delta}_0|^2\;\bar{\bf \nabla}^2 \bar{\Delta}_0\Big]
+ c|\bar{\Delta}_0|^4 \bar{\Delta}_0.
\label{eq:inhomogeneous}
\end{align}
This is a linear differential inhomogeneous equation to be solved after $\Delta_0$ is found from Eq.~(\ref{eq:GLE}). Note that similar features in the $\tau$-expansion of the gap equation appear for a two-band superconductor, as well.~\cite{shanenko}

We note again that in principle, one can continue the procedure up to arbitrary order in $\tau$, obtaining corrections to the standard GL theory with desired accuracy. While the equation for $\Delta_0$ is nonlinear, the higher order contributions to $\Delta$ will be controlled by inhomogeneous linear differential equations. Such a system of equations is solved recursively, starting from the lowest order, since solutions for previous orders will appear in the higher order equations, but not vice versa. The solution to the system will thus be uniquely defined (when the relevant boundary conditions are specified), ensuring consistency of the developed expansion.

We also remark that the structure of Eq.~(\ref{eq:GLE_corr}) makes it possible to conclude that the next-to-leading term $\Delta_1({\bf r})$ is not trivially proportional to $\Delta_0({\bf r})$. For that reason, the spatial profile of $\Delta_1({\bf r})$ is different compared to $\Delta_0({\bf r})$. This means that the characteristic length for the spatial variations of $\Delta({\bf r})$ in EGL differs from the standard GL coherence length. However, both lengths have
the same dependence on $\tau$, i.e., $\bar{\boldsymbol \nabla}\bar{\Delta}_1 \propto \tau$.

%%%%%%%%%%%%%%%%%%%%%%%%%%%%%%%%%%%%%%%%%%%%%%%%%%%%%%%%%%%%%%%%%%%%%%%%%%%%%%%%%%
\section{Series expansion in $\tau$ of the gap equation for nonzero
magnetic field}
\label{sec:tau-expansion-nonzero}
%%%%%%%%%%%%%%%%%%%%%%%%%%%%%%%%%%%%%%%%%%%%%%%%%%%%%%%%%%%%%%%%%%%%%%%%%%%%%%%%%%

The magnetic field enters the formalism developed in the previous section via changes in the normal-metal Green function. In the Gor'kov derivation the field-induced corrections are accounted for through the field-dependent Peierls phase factor as
\begin{align}
{\cal G}^{(0)}_{\rm Gor}({\bf r}t,{\bf
r}'t')=e^{\textstyle \frac{ie}{\hbar \mathfrak{c}}\int\limits_{{\bf
r}'}^{\bf r}\!{\bf A}\cdot {\rm d}{\bf q}}\;\,{\cal G}^{(0)}_{B=0}({\bf
r}t,{\bf r}'t'), \label{eq:green_B}
\end{align}
where ${\bf A}$ is the vector potential, i.e., ${\bf B}= {\rm rot}{\bf A}$. It is of importance to note that the integral in the exponent is evaluated along the straight line connecting ${\bf r}'$ and ${\bf r}$. This approximation leads to Eq.~(\ref{eq:GLE}), where the gauge-invariant derivative replaces ${\boldsymbol\nabla}$. Obtaining corrections to the GL equation requires the Green function to be calculated with an accuracy sufficient to produce the complete set of terms contributing up to the order $\tau^{5/2}$ in the $\tau$-expansion of the gap equation. Taking into account Eqs.~(\ref{eq:gap_equation_series}) and (\ref{eq:coefficients}), one concludes that the normal-metal Green function must be calculated with the accuracy ${\cal O}(\tau^2)$.

Accounting for the magnetic field in the expansion can be done by noting that the critical magnetic field $H_c$ in a superconducting system is proportional to $\tau$. Similarly a solution of the GL equations for the field also changes as $\propto \tau$. Thus the derivation of the $\tau$ expansion for the field corrections can be conveniently done by introducing the following scaling for the magnetic field:
\begin{align}
{\bf A} = \tau^{1/2} {\bar {\bf A}}, ~ {\bf B}= {\rm rot} {\bf A}= \tau {\bar {\bf B}},
\label{eq:scale_B}
\end{align}
so that the critical field quantities become independent on $\tau$ in the first order. We note that the spatial dependence of the magnetic field can also be represented in the form of the gradient expansion as it is done for the order parameter in Eq.~(\ref{eq:gradient_expansion}). As the characteristic scale for the spatial variations of the magnetic field is defined by the magnetic penetration depth $\lambda \propto \tau^{-1/2}$, which has the same $\tau$ dependence as the GL coherence length $\xi$, we can again employ the scaled spatial coordinates as those in Eq.~(\ref{eq:scaling}). So, the gradient expansion for the magnetic field follows the same rule as for the order parameter: each ${\boldsymbol\nabla}$ introduces an additional factor $\tau^{1/2}$.

The $\tau$-expansion of the Green function can now be done using the path integral method, by accounting of the quantum fluctuations around the classical trajectory. For details of the calculations we refer the reader to Appendix~\ref{sec:appendix_B}. The final expression for the Green function that contains all contributions to the order $\tau^2$ reads as
\begin{align}
&{\cal G}^{(0)}_{\omega}({\bf r},{\bf r}')=e^{{\textstyle
\frac{ie}{\hbar \mathfrak{c}}\int\limits_{{\bf r}'}^{ \bf r}\!{\bf
A}\cdot{\rm d}{\bf q}}}\left\{1+\frac{e^2\tau^2{\bar {\bf B}}^2({\bf r})}{24m^2
\mathfrak{c}^2}\right.\nonumber\\
&\times\left.\Big[\partial_{\omega}^2 +\frac{i}{\hbar} m ({\bf
r}-{\bf r}')^2_{\perp}\partial_{\omega}\Big]+{\cal O}
(\tau^{5/2})\right\}\,{\cal G}^{(0)}_{\omega,B=0}({\bf r},{\bf r}'),
\label{eq:Green_B_corr}
\end{align}
where the phase factor in the exponent is the same as in Eq.~(\ref{eq:green_B}).
We remark that the Peierls phase also contains terms of higher orders than $\tau^2$. However, they do not enter the final equations for the next-to-leading contribution to the order parameter and magnetic field. It is simply convenient to represent the Green function in the form with the Peierls factor because it naturally leads to the appearance of the gauge invariant spatial derivatives of the order parameter. One also notes that this factor is written using the unscaled quantities. In fact, the scaling does not affect it. In addition, we do not scale the difference of the coordinates ${\bf z}={\bf r}-{\bf r}'$ as it is an integration variable. The new field-dependent term in Eq.~(\ref{eq:Green_B_corr}) is proportional to $\tau^2\bar{\bf B}^2$. It does not follow from the Peierls phase and, so, is not present in the approximation given by (\ref{eq:green_B}). It is interesting that the field gradients do not contribute to the correction to the Peierls approximation.

The field-modified expansion of the gap in powers of the order parameter and its spatial derivatives are obtained by substituting Eq.~(\ref{eq:Green_B_corr}) into the kernels in Eq.~(\ref{eq:kernels}) and proceeding in a manner similar to that discussed in Appendix~\ref{sec:appendix_A} for Eq.~(\ref{eq:gap_equation_series}). It is convenient to remove the phase factor from the Green functions and introduce the ``two-point" auxiliary order parameter, i.e.,
\begin{align}
\bar{\Delta}(\bar{\bf r},\bar{\bf r}')=e^{\textstyle -\frac{2ie}{
\hbar\, \mathfrak{c}}\int\limits_{\bar{\bf r}'}^{\bar{\bf r}}\bar{
\bf A}\cdot{\rm d}\bar{\bf q}}\;\bar{\Delta}(\bar{\bf r}).
\label{eq:order_parameter_modified}
\end{align}
Then, as shown in Appendix \ref{sec:appendix_C}, the modification of Eq.~(\ref{eq:gap_equation_series}) due to the phase factor is that $\Delta({\bf r})$ is replaced by $\Delta({\bf r},{\bf r}')$, and the limit ${\bf r}' \to {\bf r}$ is implied after all relevant calculations (we remark that such a limit is not permutable with the differentiating with respect to ${\bf r}$). In particular, for the first three terms in the modified Eq.~(\ref{eq:gap_equation_series})
we have
\begin{subequations}\label{eq:nabla_shiftedA}
\begin{align}
&a_1\tau^{1/2}\lim\limits_{{\bf r}'\to{\bf r}}\bar{\Delta}
(\bar{\bf r},\bar{\bf r}')= a_1\tau^{1/2} \bar{\Delta},
\label{eq:nabla_shiftedA_a}\\
&a_2\tau^{3/2} \lim\limits_{{\bf r}'\to{\bf r}}\bar{\nabla}_{\bf
r}^2 \bar{\Delta}(\bar{\bf r},\bar{\bf r}')=a_2 \tau^{3/2}\bar{\bf
D}^2 \bar{\Delta}, \label{eq:nabla_shiftedA_b}\\
&a_3\tau^{5/2}\lim\limits_{{\bf r}'\to{\bf r}}\bar{\bf \nabla}_{\bf
r}^2 \big(\bar{\bf \nabla}_{\bf r}^2 \bar{\Delta}(\bar{\bf
r},\bar{\bf r}')\big) = \nonumber\\&\quad=a_3\tau^{5/2}\Big[\bar
{\bf D}^4-\frac{4ie}{3\hbar\,\mathfrak{c}} \bar{\rm rot}\bar{\bf
B}\cdot\bar{\bf D} + \frac{4e^2}{\hbar^2 \mathfrak{c}^2} \bar {\bf
B}^2\Big] \bar {\Delta}, \label{eq:nabla_shifted_c}
\end{align}
\end{subequations}
with $\bar {\bf D} = \bar{\boldsymbol\nabla} - i\frac{2 e}{\hbar\,\mathfrak{c}} \bar {\bf A}$, the gauge invariant gradient, and $\bar{\rm rot}\bar{\bf B}=\bar{\boldsymbol\nabla}\times \bar{\bf B}$. The terms related to the kernel $K_b$ in the field-modified
Eq.~(\ref{eq:gap_equation_series}) read
\begin{subequations}\label{eq:nabla_shiftedB}
\begin{align}
-&b_1 \tau^{3/2} \lim\limits_{{\bf r}'\to{\bf r}}|
\bar{\Delta}(\bar{\bf r},\bar{\bf r}')|^2 \bar{\Delta}(\bar{\bf
r},\bar{\bf r}')=-b_1 \tau^{3/2}
|\bar{\Delta}|^2\bar{\Delta}, \label{eq:nabla_shiftedB_a}\\
-&b_2\tau^{5/2}\lim\limits_{{\bf r}'\to{\bf r}} \Big[
2\bar{\Delta}(\bar{\bf r},\bar{\bf r}')\;|\bar{\boldsymbol\nabla}_{\bf r}
\bar{\Delta} (\bar{\bf r},\bar{\bf r}')|^2 +3
\bar{\Delta}^\ast(\bar{\bf r},\bar{\bf r}')\nonumber\\
&\;\times\big(\bar{\boldsymbol\nabla}_{\bf r}\bar{\Delta}(\bar{\bf
r},\bar{\bf r}') \big)^2 +\bar{\Delta}^2(\bar{\bf r},\bar{\bf r}')\;
\bar{\bf \nabla}_{\bf r}^2 \bar{\Delta}^{\ast}(\bar{\bf r},
\bar{\bf r}') \nonumber\\
&\;+4|\bar{\Delta}(\bar{\bf r},\bar{\bf r}')|^2 \bar{\bf
\nabla}_{\bf r}^2\bar{\Delta}(\bar{\bf r},\bar{\bf r}')\Big]=
-b_2\tau^{5/2} \Big[ 2\bar{\Delta}\;|\bar{\bf D}\bar{\Delta}|^2
\nonumber\\
&\; + 3 \bar{\Delta}^\ast (\bar{\bf D} \bar{\Delta})^2
+\bar{\Delta}^2\;(\bar{\bf D}^2\bar{\Delta})^{\ast} +4|\bar \Delta|^2
\bar{\bf D}^2\bar{\Delta}\Big], \label{eq:nabla_shiftedB_b}
\end{align}
\end{subequations}
whereas the term coming from the integral with the kernel $K_c$ is of the form
\begin{align}
c_1\tau^{5/2}\lim\limits_{{\bf r}'\to{\bf r}}|
\bar{\Delta}(\bar{\bf r},\bar{\bf r}')|^4 \bar{\Delta}(\bar{\bf
r},\bar{\bf r}')= c_1 \tau^{5/2} |\bar{\Delta}|^4\bar{\Delta}.
\label{eq:nabla_shiftedC}
\end{align}
In addition to the contributions that appear due to the Peierls factor in the Green function, we also obtain an extra contribution to the r.h.s. of  Eq.~(\ref{eq:gap_equation_series}) due to the terms proportional to ${\bf B}^2$ in Eq.~(\ref{eq:Green_B_corr}). This contribution comes only from the integral involving the kernel $K_a$~(when keeping terms up to the order $\tau^{5/2}$) and reads
\begin{align}
-a_4\tau^{5/2}\,\bar{\bf B}^2(\bar{\bf r})\lim\limits_{{\bf r}'\to
{\bf r}}\bar{\Delta}(\bar{ \bf r},\bar{\bf
r}')=-a_4\tau^{5/2}\bar{\bf B}^2\,\bar{\Delta}, \label{eq:a4}
\end{align}
where $a_4 =b\hbar^2e^2/(36m^2\mathfrak{c}^2)(1+{\cal O}(\tau))$, with $b$ given by Eq.~(\ref{eq:coefficients}).

Using the modified Eq.~(\ref{eq:gap_equation_series}) together with
Eqs.~(\ref{eq:nabla_shiftedA})-(\ref{eq:nabla_shiftedC}) and collecting terms of the same order in $\tau$, one can generalize Eqs.~(\ref{eq:GLE}) and (\ref{eq:GLE_corr}) to the case of a nonzero magnetic field. However, since the magnetic field is also a variable in the GL theory, it needs to be found self-consistently from a complementary set of equations. The most elegant way to derive the complete set of equations for the magnetic field and for the order parameter is based on the free-energy functional that accounts for the energy associated with the presence of the magnetic field and the superconducting pairing. This functional must be also represented as a series expansion in $\tau$, which is  the subject of the next section.

%%%%%%%%%%%%%%%%%%%%%%%%%%%%%%%%%%%%%%%%%%%%%%%%%%%%%%%%%%%%%%%%%%%%%%
\section{Free-energy functional}
\label{sec:free-energy}
%%%%%%%%%%%%%%%%%%%%%%%%%%%%%%%%%%%%%%%%%%%%%%%%%%%%%%%%%%%%%%%%%%%%%%

The free-energy functional ${\cal F}_s$ can be obtained, e.g., by using the path integral methods developed for the BCS theory.~\cite{popov} Its expansion in $\Delta$ reads
\begin{align}
{\cal F}_s & ={\cal  F}_{n,B=0} + \int\!\!{\rm d}^3r\,
\frac{{\bf B}^2({\bf r})}{8\pi}\,\notag \\
& +\frac{1}{g}\int\!\!{\rm d}^3r\,{\rm d}^3y\,
\big[\delta({\bf r}-{\bf y}) - K_a({\bf r},{\bf y})\big]
\Delta^*({\bf r})\Delta({\bf y})\notag\\
&-\frac{1}{2g}\int\!\!{\rm d}^3r\prod_{j=1}^3{\rm d}^3y_j \;K_b({\bf r},
\{{\bf y}\}_3)\,\Delta^{\ast}({\bf r})\Delta({\bf y}_1)\notag\\
&\times\Delta^{\ast}({\bf y}_2)\Delta ({\bf y}_3)
-\frac{1}{3g}\int\!\!{\rm d}^3r\prod_{j=1}^5 {\rm d}^3y_j \;
K_c({\bf r},\{{\bf y}\}_5) \notag \\
&\times \Delta^{\ast}({\bf r})\Delta({\bf y}_1)\Delta^{\ast}({\bf
y}_2) \Delta({\bf y}_3)\Delta^{\ast}({\bf y}_4)\Delta({\bf y}_5) -
\dots, \label{eq:functional}
\end{align}
where ${\cal F}_n$ denotes the free energy of the normal state (${\cal F}_{n,B=0}$ stays for the zero magnetic field). Here it is worth noting that, generally,
\begin{align}
K^{\ast}_a({\bf r},{\bf y})&=K_a({\bf y},{\bf r}), \nonumber\\
K^{\ast}_{b(c)}({\bf r},\{{\bf y}\}_{3(5)})&=K_{b(c)}(\{{\bf
y}\}_{3(5)},{\bf r}), \nonumber
\end{align}
which means that the r.h.s. of Eq.~(\ref{eq:functional}) is a real quantity (as necessary for the free energy). In addition, we have
\begin{align}
K_b({\bf r},{\bf y}_1,{\bf y}_2,{\bf y}_3)&=K_b({\bf y}_2,{\bf
y}_3,{\bf r},{\bf y}_1),\nonumber\\
K_c({\bf r},{\bf y}_1,{\bf y}_2,{\bf y}_3,{\bf y}_4,{\bf y}_5)&=
K_c({\bf y}_2,{\bf y}_3,{\bf r},{\bf y}_1,{\bf y}_4,{\bf y}_5)\nonumber\\
&=K_c({\bf y}_2,{\bf y}_3,{\bf y}_4,{\bf y}_5,{\bf r},{\bf y}_1),
\nonumber
\end{align}
which makes it possible to immediately find that the extremum condition of this functional with respect to $\Delta^{\ast}$ leads to Eq.~(\ref{eq:gap_perturbation}).

Series expansion of the free-energy functional given by Eq.~(\ref{eq:functional}) is obtained by following essentially a similar approach as is used in previous
sections. The scaling transformation for the order parameter, coordinates and magnetic field is introduced, and the gradient expansion for the order parameter (\ref{eq:gradient_expansion}) is substituted into Eq.~(\ref{eq:functional}). As before, the coefficients of the series expansion appears as the integrals with
kernels $K_{a,b,c}$. As we are interested in the terms that are used to derive the GL equations and the equations for the next-to-leading contributions to the order parameter and the magnetic field, we must expand the functional up to the order $\tau^3$. This follows from the rules for counting the powers of $\tau$ introduced in Sec.~\ref{sec:tau-expansion-zero}. A calculation of the series expansion in $\tau$ of the free-energy functional is greatly simplified by noting that all terms in Eq.~(\ref{eq:functional}) can be derived from the corresponding terms in Eq.~(\ref{eq:gap_equation_series}) by multiplying the latter by $(2/n)\Delta^\ast$, where $n$ is the number of $\Delta^{\ast}$-factors of the corresponding integral term in Eq.~(\ref{eq:functional}). As a result, we obtain the functional, which can be represented in the following symmetric (real) form (here and below we omit bars over the scaled quantities unless it causes confusion):
\begin{align}
f_s&=f_{n,B=0}+\frac{{\bf B}^2}{8\pi} +\frac{1}{\tau}\big(g^{-1}-a_1
\big)|\Delta|^2+ a_2 |{\bf D}\Delta|^2\notag \\
&-\tau a_3 \Big(|{\bf D}^2\Delta|^2+\frac{1}{3}
{\rm rot}{\bf B}\cdot{\bf i}+\frac{4e^2}{\hbar^2
\mathfrak{c}^2}{\bf B}^2 |\Delta|^2 \Big)
\notag \\
&+ \tau a_4 {\bf B}^2|\Delta|^2+ \frac{b_1}{2} |\Delta|^4 -\tau \frac{b_2}{2}\Big[8|\Delta|^2|{\bf D}\Delta|^2 \notag\\
&+ (\Delta^{\ast})^2 ({\bf D}\Delta)^2+\Delta^2 ({\bf D}^*\Delta^*)^2\Big]
-\tau \frac{c_1}{3}|\Delta|^6,
\label{eq:functional_real}
\end{align}
where $f_{s(n)} = {\cal F}_{s(n)}/V$~(with the scaling $f= \bar{f}\tau^2$) and ${\bf i}$ is given by
\begin{align}
{\bf i} =i\frac{2 e}{\hbar\,\mathfrak{c}}\big(\Delta {\bf D}^* \Delta^*
-\Delta^* {\bf D}\Delta\big).
\label{eq:current_i}
\end{align}
We stress that ${\bf i}$ is not the current density ${\bf j}$. However, when replacing $\Delta \to \Delta_0$ and ${\bf A} \to {\bf A}_0$ in Eq.~(\ref{eq:current_i}), we find ${\bf i}_0$ being proportional to ${\bf j}_0$, the leading contribution to the current density ${\bf j}$, see the next section.
We also note that the representation of the functional in a real symmetric form as in Eq.~(\ref{eq:functional_real}) implicitly relies on the disappearance of the corresponding surface integrals. It is easy to check that this is ensured by the boundary conditions discussed in Sec.~\ref{sec:complete-equations}.

Obtaining the final series expansion in $\tau$ for the free-energy density requires the $\tau$-expansion for the coefficients given in Eqs.~(\ref{eq:coefficients}) and (\ref{eq:a4}), for the order parameter given by Eq.~(\ref{eq:solution_expansion}), and for the magnetic field. The latter is expressed in the form
\begin{align}
{\bf A} &= {\bf  A}_0 + \tau {\bf A}_1 + \dots , ~ {\bf B} = {\bf
B}_0 + \tau {\bf B}_1 + \dots. \label{eq:field_expansion}
\end{align}
After collecting the relevant terms, the free-energy density $f_s$ is written in the form
\begin{align}
f_s - f_{n,B=0}= f_0 + \tau f_1 + {\cal O}(\tau^2),
\label{eq:functional_series_final}
\end{align}
where the leading-order term (the standard GL functional) is specified by
\begin{subequations}
\label{eq:functional_details}
\begin{align}
f_0& =\frac{{\bf B}^2_0}{8\pi}+ a|\Delta_0|^2+\frac{b}{2}|\Delta_0|^4
+{\cal K}|{\bf D}_0\Delta_0|^2,
\label{eq:energy_f0}
\end{align}
and the next-to-leading contribution $f_1$ can be written as a sum of two terms, i.e., $f_1=f_1^{(0)} + f_1^{(1)}$ with
\begin{align}
&f_1^{(0)}=\frac{a}{2}|\Delta_0|^2 + 2{\cal K}|{\bf D}_0\Delta_0|^2-{\cal Q}
\Big(|{\bf D_0}^2\Delta_0|^2 \notag \\
&+\frac{1}{3}{\rm rot}{\bf B}_0\cdot{\bf i}_0+ \frac{4e^2}{\hbar^2
\mathfrak{c}^2}{\bf B}^2_0|\Delta_0|^2\Big)
+\frac{b}{36}\,\frac{e^2\hbar^2}{m^2\mathfrak{c}^2}{\bf B}^2_0
|\Delta_0|^2\notag\\
&+b|\Delta_0|^4-\frac{{\cal L} }{2} \Big[ 8 |\Delta_0|^2 |{\bf D}_0
\Delta_0|^2 + \big(\Delta_0^{\ast}\big)^2 ({\bf D}_0 \Delta_0)^2
\notag \\
&+ \Delta_0^2 ({\bf D}^*_0 \Delta^*_0)^2\Big] - \frac{c}{3} |\Delta_0|^6
\label{eq:energy_f10}
\end{align}
and
\begin{align}
f_1^{(1)}&=\frac{{\bf B}_0\cdot {\bf B}_1}{4\pi}+\big(a+b|\Delta_0|^2\big)
(\Delta^\ast_0 \Delta_1 + \Delta_0 \Delta^\ast_1)\notag \\
&+ {\cal K}\Big[\big({\bf D}_0\Delta_0\cdot {\bf D}^{\ast}_0\Delta^{\ast}_1 +
{\rm c.c.}\big) - {\bf A}_1 \cdot {\bf i}_0\Big].
\label{eq:energy_f11}
\end{align}
\end{subequations}
Here ${\bf D}_0$ is equal to ${\bf D}$ with the substitution ${\bf A} \to {\bf A}_0$, and ${\bf i}_0$ is obtained from Eq.~(\ref{eq:current_i}) by $\Delta \to \Delta_0$ and ${\bf A} \to {\bf A}_0$.

%%%%%%%%%%%%%%%%%%%%%%%%%%%%%%%%%%%%%%%%%%%%%%%%%%%%%%%%%%%%%%%%%%%%%%%%%%%%%%%%%%%
\section{Complete set of equations for ${\bf B}\neq 0$}
\label{sec:complete-equations}
%%%%%%%%%%%%%%%%%%%%%%%%%%%%%%%%%%%%%%%%%%%%%%%%%%%%%%%%%%%%%%%%%%%%%%%%%%%%%%%%%%%

A system of coupled equations for both $\Delta$ and ${\bf A}$ are obtained from the extremum condition of the free energy given by Eq.~(\ref{eq:functional_real}). The subsequent expansion of the obtained expressions yields the GL equations as well as the equations for the next-to-leading contributions to the order parameter and the magnetic field. Alternatively, this full set of equations can also be obtained by finding the extremum of the free energy with respect to $\Delta_0$ and ${\bf A}_0$ in such a way that all terms appearing in Eq.~(\ref{eq:functional_series_final}) are taken separately. Note that finding the extremum of the functional with respect to $\Delta_1$ and ${\bf A}_1$ yields the same set of equations. This property follows from Eqs.~(\ref{eq:solution_expansion}) and (\ref{eq:field_expansion}).

By searching for the minimum of the functional with the density $f_0$, we reproduce the standard GL equations:
\begin{subequations}\label{eq:GLE_magnet_system}
\begin{align}
&a\Delta_0+b|\Delta_0|^2\Delta_0-{\cal K} {\bf D}_0^2\Delta_0=0,
\label{eq:GLE_magnet_system_a}\\
&{\rm rot} {\bf B}_0 = \frac{4\pi}{\mathfrak{c}}{\bf j}_0
\label{eq:GLE_magnet_system_b},
\end{align}
\end{subequations}
where ${\bf j}_0={\cal K} \mathfrak{c}{\bf i}_0$. Likewise, the same equations are obtained by finding the extremum of $f_1$ with respect to $\Delta_1^*$ and ${\bf A}_1$.

The minimum of the functional with the density $f_1$ with respect to $\Delta_0^*$ and ${\bf A}_0$ yields the following equations:
\begin{subequations}
\label{eq:GLE_corr_magnet}
\begin{align}
&a\Delta_1+b\big(2\,|\Delta_0|^2\Delta_1 + \Delta_0^2\Delta_1^\ast\big)
-{\cal K}{\bf D}_0^2\Delta_1=F, \label{eq:GLE_corr_magnet_a}\\
&{\rm rot}{\bf B}_1 = \frac{4\pi}{\mathfrak{c}}{\bf j}_1.
\label{eq:GLE_corr_magnet_b}
\end{align}
\end{subequations}
The r.h.s. of Eq.~(\ref{eq:GLE_corr_magnet_a}) is given by
\begin{align}
F &= -\frac{a}{2}\Delta_0+2{\cal K}{\bf D}_0^2\Delta_0+{\cal Q}
\Big[{\bf D}_0^2({\bf D}^2_0\Delta_0)\notag \\
&-i\frac{4e}{3\hbar\,\mathfrak{c}}{\rm rot} {\bf B}_0\cdot{\bf D}_0
\Delta_0+\frac{4e^2}{\hbar^2 \mathfrak{c}^2}{\bf B}_0^2 \Delta_0 \Big]
\notag \\
& - \frac{b}{36}\frac{e^2\hbar^2}{m^2\mathfrak{c}^2}{\bf
B}^2_0\Delta_0 - 2b|\Delta_0|^2\Delta_0-{\cal L}\Big[2\Delta_0|{\bf
D}_0\Delta_0|^2
\notag \\
&+3\Delta_0^\ast({\bf D}_0\Delta_0)^2+\Delta_0^2({\bf D}_0^2
\Delta_0)^\ast+4|\Delta_0|^2{\bf D}_0^2\Delta_0\Big]
\notag \\
&+c|\Delta_0|^4\Delta_0-i\frac{2e}{\hbar\,\mathfrak{c}}{\cal K}
\,\{{\bf A}_1\cdot{\bf D}_0\}_+\Delta_0,
\label{eq:F_magnet}
\end{align}
with $\{{\bf A}_1\cdot{\bf D}_0\}_+$ denoting a symmetrized product. The next-to-leading contribution to the current density ${\bf j}_1$ [${\bf j}={\bf j}_0+\tau{\bf j}_1+{\cal O}(\tau^2)$] appearing in the r.h.s. of Eq.~(\ref{eq:GLE_corr_magnet_b}) splits into two terms, i.e.,
\begin{equation}
{\bf j}_1={\cal K}\mathfrak{c}{\bf i}_1 + {\bf J},
\label{eq:current_j1}
\end{equation}
where
\begin{subequations}
\label{eq:current_j1_details}
\begin{multline}
{\bf i}_1  = i \frac{2e}{\hbar\mathfrak{c}}\big(\Delta_0 {\bf D}_0^\ast
\Delta_1^\ast + \Delta_1 {\bf D}_0^\ast \Delta_0^\ast
- \Delta_1^\ast {\bf D}_0 \Delta_0\\
- \Delta_0^\ast {\bf D}_0  \Delta_1 \big) - \frac{8 e^2}{\hbar^2
\mathfrak{c}^2} {\bf A}_1 |\Delta_0|^2,
\label{eq:i_1}
\end{multline}
and
\begin{multline}
{\bf J} = \mathfrak{c}\bigg\{\big(2{\cal K}-3{\cal L}|\Delta_0|^2\big){\bf i}_0+
{\cal Q}{\bf i}^{\prime}_0+\frac{{\cal Q}}{3}{\rm rot}~{\rm rot}~{\bf i}_0\\
+{\cal Q}\frac{8e^2}{\hbar^2\mathfrak{c}^2}\Big[{\rm rot}\big({\bf
B}_0|\Delta_0|^2\big)-\frac{1}{3}|\Delta_0|^2{\rm rot}{\bf B}_0\Big]\\
-\frac{b}{18}\frac{e^2\hbar^2}{m^2\mathfrak{c}^2}\,{\rm rot}\big({\bf
B}_0|\Delta_0|^2\big)\bigg\},\label{eq:J}
\end{multline}
with
\begin{multline}
{\bf i}^{\prime}_0 =i\frac{2 e}{\hbar \mathfrak{c}}  \Big[\Delta_0 ({\bf D}_0 {\bf D}_0^2 \Delta_0)^\ast -{\bf D}_0  \Delta_0 ({\bf D}_0^2 \Delta_0)^\ast
\\+ {\bf D}_0^2 \Delta_0 ({\bf D}_0  \Delta_0)^\ast -\Delta_0^\ast {\bf D}_0 {\bf D}_0^2 \Delta_0 \Big].
 \label{eq:i0prime}
\end{multline}
\end{subequations}
We note that, based on Eq.~(\ref{eq:GLE_magnet_system}), one finds that ${\bf i}^{\prime}_0 =(2/{\cal K})(a + b|\Delta_0|^2)\,{\bf i}_0$. We also note that ${\bf i}_1$ given by Eq.~(\ref{eq:i_1}) is the next-to-leading contribution to ${\bf i}={\bf i}_0 + \tau {\bf i}_1 + {\cal O}(\tau^2)$. Equations
(\ref{eq:GLE_corr_magnet})-(\ref{eq:current_j1_details}) are the generalization of Eq.~(\ref{eq:GLE_corr}) to the case of a nonzero magnetic field.

Through our derivation, we used the straightforward generalization of the Ginzburg-Landau boundary conditions at the specimen surface, i.e.,
\begin{align}
{\bf D}_{0\perp}\Delta_0 = 0, \quad {\bf D}_{0\perp}\Delta_1 - i \frac{2e}{\hbar\,\mathfrak{c}}{\bf A}_{1\perp}
\Delta_0 = 0. \label{eq:conditions_GL}
\end{align}
As seen, Eq.~(\ref{eq:conditions_GL}) follows from the expansion of ${\bf D}_{\perp}\Delta=0$ in $\tau$. These boundary conditions cancel the surface integrals appearing in the procedure of the variation of the free-energy functional with respect to $\Delta$ and $\Delta^{\ast}$. In addition, Eq.~(\ref{eq:conditions_GL}) allows us to cancel the surface integrals that appear due to the obvious requirement of the self-conjugation of the free-energy functional. Note that Eq.~(\ref{eq:conditions_GL}) is not the only possible choice for the boundary conditions that makes all the relevant surface integrals equal to zero. While being of great interest, discussion of different possible variants of the boundary conditions for different interfaces is beyond the scope of the present investigation. Below, based on our formulation of the EGL formalism, we investigate properties of a bulk superconductor that are not sensitive to the particular choice of the local boundary conditions at the specimen surface.

%%%%%%%%%%%%%%%%%%%%%%%%%%%%%%%%%%%%%%%%%%%%%%%%%%%%%%%%%%%%%%%%%%%%%%%%%%%%%%%%%%%
\section{Validity domain of the EGL formalism}
\label{sec:GL_domain}
%%%%%%%%%%%%%%%%%%%%%%%%%%%%%%%%%%%%%%%%%%%%%%%%%%%%%%%%%%%%%%%%%%%%%%%%%%%%%%%%%%%

%%%%%%%%%%%%%%%%%%%%%%%%%%%%%%%%%%%%%%%%%%%%%%%%%%%%%%%%%%%%%%%%%%%%%%%%%%%%%%%%%%%
\begin{figure}[t]
\resizebox{0.8\columnwidth}{!}{\rotatebox{0}{\includegraphics%
{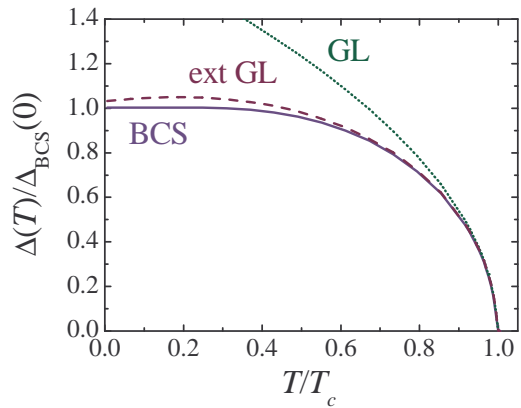}}}\caption{(Color online) The temperature dependent gap
(unscaled) in units of the zero-temperature order parameter calculated
within the full BCS approach $\Delta_{\rm BCS}(0)$ versus the relative
temperature $T/T_c$: the solid curve represents the full BCS; the
dashed curve shows the result of the EGL formalism given by Eq.~(\ref{eq:Dt_to_D0}); the dotted curve illustrates the
standard GL approach.} \label{fig1}
\end{figure}
%%%%%%%%%%%%%%%%%%%%%%%%%%%%%%%%%%%%%%%%%%%%%%%%%%%%%%%%%%%%%%%%%%%%%%%%%%%%%%%%%%%

In this section we estimate the domain of the quantitative/qualitative validity of the GL approach when extended to the next-to-leading order in $\tau$. Obviously, a detailed analysis of the accuracy of the EGL formalism, including spatially  nonuniform solutions of Eq.~(\ref{eq:GLE_corr_magnet}), requires much effort and is beyond the scope of the present work. However, it is possible to get a feeling about the accuracy of the formalism in question on the basis of the spatially uniform case. Below we compare the EGL-results for the order parameter and the thermodynamic critical field $H_c$ with those of the BCS model.

For the spatially uniform case Eq.~(\ref{eq:GLE_corr}) yields
\begin{align}
\frac{{\Delta}_1}{{\Delta}_0}\Big|_{\rm bulk} =-\frac{3}{4}-
\frac{ac}{2b^2} = -\frac{3}{4}\Big(1-\frac{31\zeta(5)}{49\zeta^2(3)}
\Big),
\label{eq:bulk}
\end{align}
with $\Delta_0=(-a/b)^{1/2}$, the solution of Eq.~(\ref{eq:GLE}). Taking into account Eq.~(\ref{eq:solution_expansion}) and using Eq.~(\ref{eq:bulk}), we obtain the order parameter in the unscaled representation up to the order $\tau^{3/2}$ as
\begin{align}
\frac{\Delta(T)}{\Delta_{\rm BCS}(0)} = e^{\gamma} \sqrt{\frac{8}{7\zeta(3)}}
\tau^{1/2} \Big[1-\frac{3}{4}\tau \Big(1-\frac{31\zeta(5)}{49\zeta^2(3)}\Big)\Big],
\label{eq:Dt_to_D0}
\end{align}
where $\Delta_{\rm BCS}(0)=(\pi/e^{\gamma})T_c$ is the zero-temperature gap calculated from the full BCS formalism, see, e.g., Ref.~\onlinecite{fett}. Results found from the standard and extended GL approaches are compared to the full BCS solution in Fig.~\ref{fig1}. We can see that the GL result notably differs from the BCS curve below temperatures $T=0.7$-$0.8T_c$. At the same time the EGL approach is in a very good quantitative agreement with the BCS theory down to $T=0.2T_c$, and only below this temperature the order parameter calculated within the extended formalism exhibits a slight decrease not supported by the BCS picture.

\begin{figure}[t]
\resizebox{0.8\columnwidth}{!}{\rotatebox{0}{\includegraphics{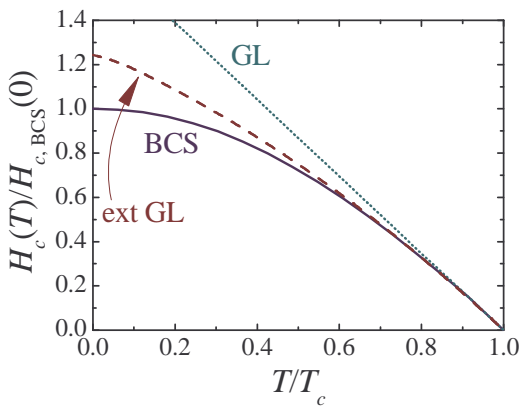}}}
\caption{(Color online) The thermodynamic critical magnetic field $H_c(T)$~(unscaled) in units of $H_{c,{\rm BCS}}(0)$ versus the relative temperature $T/T_c$: the solid curve represents the BCS theory; the dashed curve shows the result of the EGL formalism; the dotted curve illustrates the standard GL theory.} \label{fig2}
\end{figure}

The thermodynamic critical field $H_c$ measures the condensation energy so that
\begin{align}
\frac{H_c^2}{8\pi} = f_{n,B=0} - f_{s,B=0},
\label{eq:Hc}
\end{align}
where $f_{s,B=0}$ is the free-energy density of a homogeneous superconducting state in the absence of a magnetic field. Using Eqs.~(\ref{eq:GLE}), (\ref{eq:GLE_corr}), (\ref{eq:functional_series_final}) and (\ref{eq:functional_details}), we find
\begin{align}
& H_c= H_{c\,0} + \tau H_{c1}+{\cal O}(\tau^2), \notag \\
& H_{c\,0} = \sqrt{\frac{4\pi a^2}{b}},\quad H_{c1}=-H_{c\,0}
\left(\frac{1}{2}+\frac{ac}{3b^2}\right).
\label{eq:critical_field}
\end{align}
The solution $H_c=H_{c\,0}$ recovers the result of the GL theory~($H_c = \tau H_{c0}$ in the original unscaled variables). The term $\tau H_{c1}$ ($\tau^2 H_{c1}$ in the original unscaled variables) provides the next order correction.
The numerical coefficient $ac/(3b^2)$ is calculated using Eq.~(\ref{eq:coefficients}).
This yields
\begin{align}
H_c/H_{c\,0} = 1-\frac{\tau}{2}\Big(1-\frac{31\zeta(5)}{49\zeta^2(3)}
\Big)+{\cal O}(\tau^2)&\nonumber\\
= 1-0.273\tau +{\cal O}(\tau^2)&. \label{eq:critical_field_A}
\end{align}
Being back to the original unscaled variables, we get for the thermodynamic critical field up to the order $\tau^2$
\begin{equation}
\frac{H_c(T)}{H_{c,{\rm BCS}}(0)}= e^{\gamma} \sqrt{\frac{8}{7\zeta(3)}} \tau
\Big[1-\frac{\tau}{2}\Big(1-\frac{31\zeta(5)}{49\zeta^2(3)}
\Big)\Big],
\label{eq:Ht_to_HBCS0}
\end{equation}
where $H_{c,{\rm BCS}}(0)=[4\pi N(0)]^{1/2}\,\pi T_c/e^{\gamma}$ is the zero-temperature thermodynamic critical field.~\cite{fett} Figure~\ref{fig2} shows the result given by Eq.~(\ref{eq:Ht_to_HBCS0}) as compared to those of the standard GL formalism and the BCS approach. Here one notes a very good quantitative agreement between the EGL formalism and the BCS theory down to temperatures $0.3$-$0.4T_c$. In particular, the EGL results are only by $5\%$ larger at $T=0.35T_c$. At lower temperatures the curve representing the EGL formalism deviates from the BCS data by $10$-$20\%$. This can be compared with the standard GL theory for which $H_c(0)/H_{c,{\rm BCS}}= e^{\gamma} \sqrt{8/(7\zeta(3))} =1.736$.

%%%%%%%%%%%%%%%%%%%%%%%%%%%%%%%%%%%%%%%%%%%%%%%%%%%%%%%%%%%%%%%%%%
%%%%%%%%%%%%%%%%%%%%%%%%%%%%%%%%%%%%%%%%%%%%%%%%%%%%%%%%%%%%%%%%%%%%%%%%%%%%%%%%%%%

Thus, based on the results given in Figs.~\ref{fig1} and \ref{fig2}, one is able to expect that the domain of the quantitative validity of the EGL theory (in the clean limit) is $\tau < 0.7\,(T/T_c > 0.3)$, which is a significant extension as compared to $\tau < 0.2$-$0.3$, typical for the standard GL approach.

%%%%%%%%%%%%%%%%%%%%%%%%%%%%%%%%%%%%%%%%%%%%%%%%%%%%%%%%%%%%%%%%%%%%%%%%%%%%%%%%%%%
\section{Surface energy in the next-to-leading order in $\tau$}
\label{sec:surface-energy}
%%%%%%%%%%%%%%%%%%%%%%%%%%%%%%%%%%%%%%%%%%%%%%%%%%%%%%%%%%%%%%%%%%%%%%%%%%%%%%%%%

The EGL formalism given by Eq.~(\ref{eq:GLE_corr_magnet}) allows one to
find corrections to the solutions of the physical problems for which the GL approach is relevant. Here we illustrate the power of the EGL formalism by investigating, in the next-to-leading order in $\tau$, the energy associated with a surface separating the superconducting and normal phases. It is one of the fundamental problems in the theory of superconductivity in view of the fact that superconducting materials are classified as type I or type II according to whether the surface energy is positive or negative, respectively. From the standard GL theory it is well-known that the surface energy is controlled by the GL parameter $\kappa = \lambda/\xi$, where $\lambda$ is the magnetic penetration depth and $\xi$ is the GL coherence length. The surface energy is positive for $\kappa < \kappa^*$ and negative for $\kappa > \kappa^*$, where $\kappa^*= 1/\sqrt{2}$ is a universal constant, being independent of temperature. Now, based on the EGL approach, it is interesting to check whether or not $\kappa^*$ is independent of temperature in the next-to-leading order in $\tau$.

Following the standard calculation of the surface energy, see, e.g., Ref.~\onlinecite{Landau9}, we consider a surface separating the superconducting and normal phases perpendicular to the $z$-axis. The superconducting and normal phases
are found at $z\to -\infty$ and $z\to \infty$, respectively. This means that the superconducting order parameter approaches its bulk (uniform) value at $z \to -\infty$ and goes to zero at $z\to \infty$. The magnetic field is chosen in the $y$-direction. It approaches the thermodynamic critical field $H_c$~[see Eq.~(\ref{eq:critical_field})] at $z\to \infty$ and becomes zero at $z\to -\infty$. Note that $z\to\pm\infty$ means here that the point is far beyond the surface but still inside the specimen. When going far beyond the specimen, the magnetic field always approaches $H_c$. Further, the vector potential is taken in the Coulomb gauge in the form ${\bf A}= \{A(z),0,0\}$. The magnetic field respectively reads as ${\bf B} = \{0,A^{\prime}(z),0\}$. Hereafter the prime sign denotes the derivative with respect to $z$. It should be noted that this choice applies to both ${\bf B}_0$ (${\bf A}_0$) and ${\bf B}_1$ (${\bf A}_1$). In this case Eqs.~(\ref{eq:GLE_magnet_system_a}) and (\ref{eq:GLE_corr_magnet_a}) contain only real coefficients and, hence, have real solutions.

For our choice the surface energy per unit area $\sigma_{sn}$ (the surface tension) is given by a 1D integral of the difference between the Gibbs free-energy densities of the nonuniform superconducting solution $g_s$ and the uniform normal state $g_n$, i.e.,
\begin{align}
\sigma_{sn} =\int\limits_{-\infty}^{+\infty}\!\!{\rm d}z\,(g_s -
g_n), \quad  g_{s(n)}= f_{s(n)} - \frac{\bf H B}{4\pi},
\label{eq:energy_modified}
\end{align}
where $f_s$ is given by Eqs.~(\ref{eq:functional_series_final}) and (\ref{eq:functional_details}) and $f_n=f_{n,B=0} + {\bf H}^2/(8\pi)$. The external magnetic field ${\bf H}$ is constant and its absolute value is equal to the thermodynamic critical field $H_c$. In the normal phase ${\bf B}={\bf H}$ and $B=H=H_c$, which results in $g_n = f_{n,B=0} - H_c^2/(8\pi)$. To proceed further, it is convenient to introduce the following dimensionless quantities:
\begin{align}
&\widetilde{{\bf r}}=\frac{{\bf r}}{\lambda},\,\widetilde{{\bf A}}=
\frac{{\bf A}}{H_{c\,0}\lambda},\,\widetilde{{\bf B}}=\frac{{\bf B}}{H_{c\,0}},
\widetilde{\Delta}=\sqrt{-\frac{b}{a}}\,\Delta,
\nonumber\\
&\widetilde{{\bf i}}_0={\bf i}_0 \frac{4\pi{\cal K}
\lambda}{H_{c\,0}},\,\widetilde{g}_{s(n)}=\frac{4\pi}{H^2_{c\,0}} g_{s(n)},
\,\widetilde{\sigma}_{sn} = \frac{4\pi}{\lambda H^2_{c\,0}} \sigma_{sn},
\label{eq:scaling_dimensionless}
\end{align}
with $\lambda=(\hbar \mathfrak{c}/|e|) \sqrt{-b/(32\pi{\cal K}a)}$. In what follows, we omit the tilde in all the formulas unless it causes confusion. Using the new dimensionless quantities, we arrive at
\begin{subequations}
\label{eq:surf_tens_expansion}
\begin{align}
\label{eq:surface_tension} &\sigma_{sn}=\alpha_0 + \tau\alpha_1,
\quad\alpha_i = \int\limits_{-\infty}^{+\infty}\!\!{\rm d}z\,
g_i(z),
\end{align}
where $g_0(z)$ and $g_1(z)$ are the coefficients of the expansion of the difference $g_s-g_n$ in powers of $\tau$, i.e., $g_s-g_n=g_0+g_1\tau + {\cal O}(\tau^2)$. Taking into account Eqs.~(\ref{eq:energy_f0}) and (\ref{eq:energy_modified}), we obtain
\begin{align}
g_0=&-\frac{\Delta_0\Delta^{\prime\prime}_0}{\kappa^2}+\left(\frac{A_0^2}{2}
-1 \right) \Delta^2_0 +\frac{\Delta^4_0}{2}
+\frac{1}{2} (A^{\prime}_0 - 1 )^2,
\label{eq:g0}
\end{align}
From Eqs.~(\ref{eq:energy_f10}), (\ref{eq:energy_f11}) and  (\ref{eq:energy_modified}), it is seen that $g_1$ can be split into two parts, i.e.,
\begin{align}
g_1 = g_1^{(0)} + g_1^{(1)},
\label{eq:g2}
\end{align}
where $g^{(0)}_1$ includes only the quantities with the index $0$, i.e.,
\begin{align}
g^{(0)}_1 =&-\frac{\Delta^2_0}{2} + 2\frac{(\Delta^{\prime}_0)^2}{\kappa^2}+
A^2_0\Delta^2_0 +\frac{{\cal Q}a}{{\cal K}^2}
\Big[\Big(\frac{\Delta^{\prime\prime}_0}{\kappa^2}\! -
\!\frac{A^2_0}{2} \Delta_0\Big)^2\nonumber\\
&\;\; + \frac{1}{3\kappa^2}A_0^{\prime\prime}A_0\Delta_0^2+
\frac{(A^{\prime}_0)^2}{2\kappa^2}\Delta^2_0\Big]+ \frac{(A^{\prime}_0)^2 \Delta^2_0}{48(k_F\lambda)^2}
+\Delta^4_0\nonumber\\ &\;\;+ \frac{{\cal
L}a}{2b{\cal K}}\Big[\frac{10}{\kappa^2}(\Delta_0^{\prime})^2\Delta_0^2  +
3A_0^2\Delta_0^4\Big]+\frac{ac}{3b^2}\Delta^6_0,\label{eq:g10}
\end{align}
and $g^{(1)}_1$ reads
\begin{align}
g_1^{(1)} = &2\Big[(\Delta^2_0-1)\Delta_0\Delta_1+ \frac{A^2_0}{2}\Delta_0\Delta_1+\frac{1}{\kappa^2}\Delta^{\prime}_0 \Delta^{\prime}_1\Big]
\nonumber\\
&-A_1 i_0 + (A^{\prime}_0-1)\Big(A_1^{\prime}+\frac{1}{2}+\frac{ac}{3b^2}\Big),
\label{eq:g11}
\end{align}
\end{subequations}
incorporating $\Delta_1$ and $A_1$.

When calculating $\sigma_{sn}$, we use $\Delta_{0(1)}$ and ${\bf A}_{0(1)}$       obtained by solving Eqs.~(\ref{eq:GLE_magnet_system}) and (\ref{eq:GLE_corr_magnet}). This makes it possible to simplify the problem significantly, excluding the terms involving $\Delta_1$ and $A_1$. The point is that such terms in $g^{(1)}_1$ can be transformed into the functional derivatives of $f_0$~[see Eq.~(\ref{eq:energy_f0})] with respect to $\Delta_0$ and $A_0$. These derivatives are equal to zero and generate the standard GL equations (\ref{eq:GLE_magnet_system_a}) and (\ref{eq:GLE_magnet_system_b}) that are now reduced to
\begin{subequations}
\label{eq:GLE_scaled}
\begin{align}
-&\frac{1}{\kappa^2}\Delta_0^{\prime \prime} - \Delta_0
+\Delta_0^3+\frac{A_0^2}{2}\Delta_0= 0,
\label{eq:GLE_scaled_1}\\
& A_0^{\prime \prime}  = -i_0 = A_0 \Delta_0^2,
\label{eq:GLE_scaled_2}
\end{align}
with the boundary conditions (inside the sample)
\begin{align}
&\Delta_0(-\infty) = 1, ~ A_0^\prime(-\infty) = 0, \nonumber\\
&\Delta_0(\infty) = 0, ~ A_0^\prime(\infty) = 1.
\label{eq:GLE_scaled_boundary}
\end{align}
\end{subequations}
Here we stress again that far outside the specimen we always have $A'_0 = 1$ and $A'_1= -\frac{1}{2}-\frac{ac}{3b^2}$. However, inside the sample, deep in the superconducting domain, we are able to employ $A'_{0(1)}=0$~(in addition, due to our choice of a real order parameter, we also have $A_{0(1)}=0$). So, based on Eqs.~(\ref{eq:surf_tens_expansion}) and (\ref{eq:GLE_scaled}), we find that $g^{(1)}_1$ can be replaced by
\begin{align}
g_1^{(1)} =\Big(\frac{1}{2}+\frac{ac}{3b^2}\Big)\big(A^{\prime}_0-
1\big).
\end{align}
As seen, there are only two physical parameters that enters the relevant expressions for $\sigma_{sn}$, i.e., the GL parameter $\kappa$ and the product of the Fermi wavenumber and the magnetic penetration depth $k_F\lambda$. As to the quantities ${\cal Q}a/{\cal K}^2$, $ac/(3b^2)$, ${\cal L}a/(b{\cal K})$ and $H_{c1}/H_{c\,0}$, they are simply numbers. From Eq.~(\ref{eq:coefficients})
we obtain
\begin{align}
\frac{{\cal Q}a}{{\cal K}^2}=-0.817,\,\frac{ac}{3b^2}=-0.227,\,
\frac{{\cal L}a}{b{\cal K}}=-0.454
\label{eq:numbers}
\end{align}
and, in addition, $H_{c1}/H_{c\,0}=-0.273$, as seen from
Eqs.~(\ref{eq:critical_field}) and (\ref{eq:critical_field_A}). For conventional superconductors the term including $k_F\lambda$ in Eq.~(\ref{eq:g10}) is extremely small and, so, we are left with only one governing physical parameter $\kappa$
in both the leading and next-to-leading orders.

Now, we have everything at our disposal to calculate $\kappa^*$ at which the surface energy $\sigma_{sn}$ becomes zero, i.e.,
\begin{align}
\alpha_0(\kappa^*) + \tau \alpha_1(\kappa^*) = 0.
\label{eq:kappa_equation}
\end{align}
The solution to Eq.~(\ref{eq:kappa_equation}) should be represented in the form of the $\tau$-expansion
\begin{align}
\kappa^* = \kappa^*_0 + \tau \kappa^*_1 + {\cal O}(\tau^2),
\label{eq:kappa}
\end{align}
where $\kappa^*_0$ obeys
\begin{align}
\alpha_0(\kappa^*_0) = 0.
\label{eq:kappa0}
\end{align}
When $\kappa^*_0$ is known, the next-to-leading order contribution $\kappa^*_1$ is found by expanding Eq.~(\ref{eq:kappa_equation}) in
powers of $\tau$, i.e.,
\begin{align}
\alpha_0(\kappa^*_0)+&\tau \alpha_1(\kappa^*_0)+\tau\kappa^*_1
\bigg[\frac{\partial\alpha_0}{\partial \kappa^*_0}
\notag \\
&\;+\int\limits_{-\infty}^{+\infty}\!\!\!{\rm d}z\,\Big(\frac{\delta
\alpha_0}{\delta \Delta_0}\,\frac{\partial\Delta_0}{\partial \kappa^*_0}
+\frac{\delta\alpha_0}{\delta A_0}\,\frac{\partial A_0}{\partial\kappa^*_0}  \Big)\bigg]=0,
\label{eq:functional_derivatives}
\end{align}
where all contributions are calculated at $\kappa = \kappa^*_0$. The partial derivative $\frac{\partial \alpha_0}{\partial \kappa^*_0}$ in the above expression accounts for a change in $\alpha_0$ caused by the explicit presence of $\kappa$ in Eq.~(\ref{eq:GLE_scaled_1}). The functional derivatives of $\alpha_0$ appear in the integral in the r.h.s. of Eq.~(\ref{eq:functional_derivatives}) due to changes in $\Delta_0$ and $A_0$ when changing $\kappa$. At first sight, this complicates the problem enormously because of the need to know the derivatives $\frac{\partial\Delta_0}{\partial \kappa}$ and $\frac{\partial A_0}{\partial \kappa}$ at the point $\kappa=\kappa^*_0$. It is, however, clear that $\frac{\delta \alpha_0}{\delta \Delta_0}=0$ and $\frac{\delta \alpha_0}{\delta A_0}=0$ because they yield the GL equations in the leading order in $\tau$. So, $\kappa^*_1$ is given by
\begin{align}
\kappa_1 = - \alpha_1(\kappa^*_0)\left(\frac{\partial \alpha_0}{\partial \kappa^*_0}\right)^{-1},
\label{eq:kappa_1}
\end{align}
and the derivatives of $\Delta_0$ and $A_0$ with respect to $\kappa$ at the point $\kappa=\kappa^*_0$ disappear from the final result.

Thus, to calculate both $\kappa_0$ and $\kappa_1$, one only needs to find $\Delta_0$ and $A_0$ from the standard GL equations (\ref{eq:GLE_scaled_1}) and
(\ref{eq:GLE_scaled_2}) at $\kappa =\kappa^*_0$. The calculation of $\kappa^*_0=1/\sqrt{2}$ is a classical problem that can be found in textbooks on superconductivity, see, e.g., Ref.~\onlinecite{Landau9}. Calculating $\kappa_1$, we first find
\begin{align}
&\frac{\partial \alpha_0}{\partial \kappa^*_0} =
-\frac{1}{2(\kappa^*_0)^3} {\cal I}_1, \quad {\cal I}_1 =
\int\limits_{-\infty}^{+\infty}\!\!\!{\rm d}z\,\Delta_0^2
\big(1- \Delta_0^2\big),
\label{eq:alpha0_div}
\end{align}
where Eq.~(\ref{eq:GLE_scaled_1}) was used to simplify the expression. In turn, using Eqs.~(\ref{eq:surf_tens_expansion}), (\ref{eq:GLE_scaled}) and the helpful relation~\cite{Landau9}
\begin{equation}
\Delta^2_0=1-A'_0
\label{eq:delta0_A0}
\end{equation}
and assuming $k_F\lambda \gg 1$, which is always satisfied in the conventional superconductors, one finds
\begin{multline}
\alpha_1={\cal I}_1\Big(1+\frac{2{\cal Q}a}{{\cal K}^2}-\frac{ac}{3b^2}\Big)\\
+{\cal I}_2 \Big(\frac{2{\cal L}a}{b{\cal K}}-\frac{5{\cal Q}a}{3 {\cal K}^2}- \frac{ac}{3b^2}\Big),
\label{eq:alpha1}
\end{multline}
with
\begin{align}
{\cal I}_2 = \int\limits_{-\infty}^{+\infty}\!\!{\rm d}z\,\Delta_0^4\big(1-
\Delta_0^2\big).
\end{align}
Numerically solving Eqs.~(\ref{eq:GLE_scaled}), we obtain ${\cal I}_1=0.775$ and ${\cal I}_2 =0.433$. Finally, substituting Eqs.~(\ref{eq:alpha0_div})
and (\ref{eq:alpha1}) into Eq. (\ref{eq:kappa_1}) and using Eq.~(\ref{eq:numbers}), we obtain $\kappa^*_1=-0.027\kappa^*_0$. Thus, Eq.~(\ref{eq:kappa}) for $\kappa^*$ reads
\begin{equation}
\kappa^* = \frac{1}{\sqrt{2}}\big(1 - 0.027\tau +{\cal O}(\tau^2)\big).
\label{eq:kappa_final}
\end{equation}
As seen, contrary to the result of the standard GL theory~(the leading order in our $\tau$-expansion), $\kappa^*$ given by Eq.~(\ref{eq:kappa_final}) is temperature-dependent. This means that the traditional classification of type-I and type-II superconductors becomes, in principle, temperature-dependent. However, inconvenience caused by such dependence is not crucial because of the very small value of $\kappa^*_1$, i.e., a change in $\kappa^*$ with temperature is less than $2\%$ in the strict validity domain of the EGL theory.

%%%%%%%%%%%%%%%%%%%%%%%%%%%%%%%%%%%%%%%%%%%%%%%%%%%%%%%%%%%%%%%%%%%%%%%%%%%%%%%%%%%
\section{Conclusions}
\label{sec:conclusions}
%%%%%%%%%%%%%%%%%%%%%%%%%%%%%%%%%%%%%%%%%%%%%%%%%%%%%%%%%%%%%%%%%%%%%%%%%%%%%%%%%%%

Employing an approach similar to the asymptotic-expansion methods used in the mathematical physics, we constructed a systematic expansion of the self-consistent gap equation (for a clean s-wave single-band superconductor) in powers of $\tau$, the proximity to the critical temperature. The procedure of matching the expansion terms of the same order of magnitude generates a hierarchy of equations. The lowest-order theory, i.e., the equations for the leading contributions to the order parameter and the magnetic field, recovers the standard GL approach, while the next orders in $\tau$ constitute its extension. Such a hierarchy of equations should be solved recursively, starting from the standard GL equations. We derived and studied the equations for the next-to-leading contributions to the order parameter $\Delta$ and the magnetic field ${\bf B}$. In order to select all relevant terms in the case of a nonzero magnetic field, the normal-metal temperature Green function was generalized beyond the standard Peierls phase approximation to incorporate additional terms up to the order $\tau^2$. The relevant boundary conditions were shown to be directly related to the series expansion in $\tau$ for the current. The accuracy of the GL theory extended to the next-to-leading order in $\tau$ was tested by comparing the results of the extended formalism for the uniform order parameter and the critical magnetic field with the corresponding results of the standard GL approach and the BCS theory. This demonstrated that the validity domain of the GL theory is considerably increased by the extension. We found very good agreement with the full BCS calculations down to temperatures $0.3$-$0.4\,T_c$. To illustrate advantages of the constructed extension to the GL formalism, the surface energy for the interface between the superconducting and normal phase was investigated. We have found, in a semi-analytical form, the temperature-dependent correction to the value of the GL parameter $\kappa$ at which the surface energy becomes zero. Surprisingly, the obtained correction is extremely small: it does not exceed $2\%$ even at $T=0.3$-$0.4\,T_c$. This result implies that the boundary between type-I and type-II superconducting behavior is almost independent of temperature.

It should be noted that a functional similar to Eq.~(\ref{eq:functional_real}) was considered in Ref.~\onlinecite{mineev} in the context of the FFLO state (we have an additional term $\propto a_4$). However, there is a conceptional difference with our work: we focus on a series expansion in $\tau$ and, so, this functional is only the initial point to construct such an expansion. Our focus on a perturbation theory in $\tau$ allowed us to make a proper selection of all the relevant contributions. It means that we did not simply borrow some functional from previous papers as the initial step for our study but instead performed an extensive procedure of microscopic derivations (see Appendices) accompanied by an accurate analysis of the temperature dependence of each contributing term. We proved that the initial free-energy functional given by Eq.~(\ref{eq:functional_real}) contains all the terms that contribute to $\Delta$ and ${\bf B}$ in the leading and next-to-leading orders in $\tau$.

We also note that though the term $\propto a_4$ in Eq.~(\ref{eq:functional_real}) does not produce a pronounced contribution for conventional bulk superconductors [it is proportional to $1/(k_F\lambda)^2$, see, e.g., Eq.~(\ref{eq:g10})], going beyond the Peierls phase approximation can be of importance, e.g., for multi-band (subband) materials/systems, where one of the relevant bands is characterized by an extremely small Fermi momentum but $\lambda$ is determined mostly by other bands with large $k_F$~(so that the product $k_F\lambda$ can be even smaller than $1$). One of possible examples is single-crystalline metallic superconducting nanofilms, where different subbands are induced by the quantization of the perpendicular electron motion and the energetic position of each subband (with respect to the Fermi level) can vary significantly with changing nanofilm thickness, substrate material etc.~\cite{chen} Similar physics can be expected for a thin superconducting layer induced by an external electric field at the interface between the semiconductor (${\rm SrTiO_3}$ and ${\rm KTaO_3}$) and an electrolyte.~\cite{ueno}

\begin{acknowledgments}
This work was supported by the Flemish Science Foundation (FWO-Vl) and
the Belgian Science Policy (IAP). A.V.V. is grateful to V. Zalipaev for important comments. A.A.S. thanks W. Pogosov for helpful notes. Discussions with E.~H.~Brandt and A.~Perali are appreciated.
\end{acknowledgments}

\appendix

%%%%%%%%%%%%%%%%%%%%%%%%%%%%%%%%%%%%%%%%%%%%%%%%%%%%%%%%%%%%%%%%%%%%%%%%%%%%%%%%%%%
\section{Coefficients in the expansion of the gap equation for zero magnetic field}
\label{sec:appendix_A}
%%%%%%%%%%%%%%%%%%%%%%%%%%%%%%%%%%%%%%%%%%%%%%%%%%%%%%%%%%%%%%%%%%%%%%%%%%%%%%%%%%%

%%%%%%%%%%%%%%%%%%%%%%%%%%%%%%%%%%%%%%%%%%%%%%%%%%%%%%%%%%%%%%%%%%%%%%%%%%%%%%%%%%%
\subsection{Coefficients $a_i$ related to the integral kernel $K_a({\bf r},{\bf y})$}
\label{subsec:A1}
%%%%%%%%%%%%%%%%%%%%%%%%%%%%%%%%%%%%%%%%%%%%%%%%%%%%%%%%%%%%%%%%%%%%%%%%%%%%%%%%%%%

We start our derivation of the r.h.s. of Eq.~(\ref{eq:gap_equation_series}) from the terms coming from the integral involving the kernel $K_a({\bf r},{\bf y}) = K_a({\bf z})$~(with ${\bf z}={\bf r}-{\bf y}$), i.e.,
\begin{equation}
I_a=\frac{1}{g}\int\!\!{\rm d}^3z\, K_a({\bf z})\Delta({\bf r}+{\bf z}).
\label{eq:int_Ka}
\end{equation}
Following the usual practice, this integral is expanded in terms of the spatial derivatives of the order parameter $\Delta({\bf r})$, i.e., Eq.~(\ref{eq:gradient_expansion}). Keeping to the mnemonic rule formulated in Sec.~\ref{sec:tau-expansion-zero}, we conclude that working to the order $\tau^{5/2}$, it is necessary to incorporate all the spatial derivatives up to the fourth order in the gradient expansion Eq.~(\ref{eq:gradient_expansion}). Due to the symmetry of the kernel $K_a({\bf z})$ with respect to the transformation ${\bf z} \to -{\bf z}$, the first- and third-order derivatives do not contribute. So, we obtain only the three relevant terms in Eq.~(\ref{eq:gap_equation_series}) that come from Eq.~(\ref{eq:int_Ka}), i.e.,
\begin{subequations}\label{eq:Ia}
\begin{align}
&I^{(1)}_a=\frac{\tau^{1/2}}{g}\bar{\Delta}\int\!\!{\rm d}^3z\,
K_a({\bf z}),\label{eq:Ia_1}\\
&I^{(2)}_a=\frac{\tau^{3/2}}{g}\int\!\!{\rm d}^3z\, K_a({\bf z})
\frac{({\bf z}\cdot\bar{{\boldsymbol\nabla}})^2}{2!}\bar{\Delta},
\label{eq:Ia_2}\\
&I^{(3)}_a=\frac{\tau^{5/2}}{g}\int\!\!{\rm d}^3z\, K_a({\bf z})
\frac{({\bf z}\cdot\bar{{\boldsymbol\nabla}})^4}{4!}\bar{\Delta},
\label{eq:Ia_3}
\end{align}
\end{subequations}
where $\bar{\Delta}(\bar{\bf r})$ is the scaled order parameter as a function of scaled coordinates. Details on calculations of $I^{(1)}_1$ and $I^{(2)}_a$ can be found in textbooks, e.g., in Ref.~\onlinecite{fett}, with the result
\begin{align}
I^{(1)}_a = a_1\tau^{1/2} \bar{\Delta}, \quad I^{(2)}_a = a_2 \tau^{3/2} \bar{\nabla}^2\bar{\Delta},
\label{eq:Ia_12a}
\end{align}
where $a_1$ and $a_2$ are given by Eq.~(\ref{eq:coefficients}). To find $I^{(3)}_a$, it is first convenient to rearrange Eq.~(\ref{eq:Ia_3a}) in the form (odd powers of $\bar{\nabla}$ do not contribute)
\begin{multline}
I^{(3)}_a=\frac{\tau^{5/2}}{g}\int\!\!{\rm d}^3z\, K_a({\bf z})
\Big[\frac{1}{4!}\sum\limits_n z^4_n \bar{\nabla}^4_n\\
+\frac{1}{8}\sum\limits_{n\not=m}z^2_n z^2_m\bar{\nabla}^2_n
\bar{\nabla}^2_m\Big]\,\bar{\Delta},
\label{eq:Ia_3a}
\end{multline}
with ${\bf z}=\{z_1,z_2,z_3\}$ and $\bar{\boldsymbol\nabla} =
\{\bar{\nabla}_1,\bar{\nabla}_2,\bar{\nabla}_3\}$. As seen from
Eq.~(\ref{eq:Ia_3a}), two integrals are needed to be calculated,
i.e.,
$$
\int\!\!{\rm d}^3z\;K_a({\bf z})\,z^4_n\;\;{\rm and}\;\;
\int\!\!{\rm d}^3z\;K_a({\bf z})\,z^2_n z^2_m \;\;(n\not=m).
$$
Below we assume spherical symmetry, i.e., $K_a({\bf z})=K_a(z)$,
with $z\equiv|{\bf z}|$. In this case the above integrals do not
depend on indices $n$ and $m$.

As follows from Eq.~(\ref{eq:green_0}),
\begin{equation}
{\cal G}^{(0)}_{\omega}({\bf r},{\bf y})= -\frac{\pi N(0)}{k_F |{\bf
r}-{\bf y}|}\;e^{i\,k_F|{\bf r}-{\bf y}|\,{\rm sgn}\omega
-i\frac{|\omega|}{v_F}|{\bf r}-{\bf y}|}, \label{eq:Green_func_coor}
\end{equation}
with $k_F$ the Fermi wavenumber, $N(0)=mk_F/(2\pi^2\hbar^2)$ the
density of states at the Fermi surface and $v_F$ the Fermi velocity.
When inserting Eq.~(\ref{eq:Green_func_coor}) into Eq.~(\ref{eq:kernels}),
one can find [${\widetilde {\cal G}}^{(0)}_{\omega}({\bf r},{\bf y})=
-{\cal G}^{(0)}_{-\omega}({\bf y},{\bf r})$]
\begin{equation}
K_a({\bf z})=gT\Big[\frac{\pi N(0)}{k_F z}\Big]^2 \; \frac{1}{{\rm
sinh}\big(z/\xi_T\big)}, \label{eq:Ka_coor}
\end{equation}
with $\xi_T=\hbar v_F/(2\pi T)$. Then, based on Eq.~(\ref{eq:Ka_coor})
and replacing $T$ by $T_c$, we obtain
\begin{subequations}\label{integrals}
\begin{align}
&\int\!\!{\rm d}^3z\;K_a({\bf z})\,z^4_n=\frac{4}{5}\,gc \,\hbar^4v^4_F,\\
&\int\!\!{\rm d}^3z\;K_a({\bf z})\,z^2_n z^2_m=\frac{4}{15}\,g
c\,\hbar^4 v^4_F \; \;\;(n\not=m),
\end{align}
\end{subequations}
where $c$ is given by Eq.~(\ref{eq:coefficients}). The integrals can
easily be taken by using (see, e.g., Appendixes in Ref.~\onlinecite{fett})
$$
\int\limits_0^{\infty}\!\!{\rm d}x \;\frac{x^{\ell-1}}{{\rm sinh}(x)} =
2(1-2^{-\ell})\Gamma(\ell) \zeta(\ell)\quad (\ell >1),
$$
with $\Gamma(\ldots)$ the Euler gamma-function and $\zeta(\ldots)$
the Riemann zeta-function. Equations (\ref{eq:Ia_3a}) and
(\ref{integrals}) make it possible to get
\begin{align}
I^{(3)}_a = \tau^{5/2} \frac{c}{30}\hbar^4 v^4_F\Big[\sum\limits_n
\bar{\nabla}^4_n
+\sum\limits_{n\not=m}\bar{\nabla}^2_n\bar{\nabla}^2_m
\Big]\;\bar{\Delta}, \label{eq:Ia_3b}
\end{align}
which is reduced to
\begin{align}
I^{(3)}_a = a_3\tau^{5/2}\bar{\nabla}^2(\bar{\nabla}^2\bar{\Delta}),
\label{eq:Ia_3c}
\end{align}
with the coefficient $a_3$ given by Eq.~(\ref{eq:coefficients}).
Now, collecting the results for $I^{(1)}_a$, $I^{(2)}_a$ and
$I^{(3)}_a$, we obtain
\begin{align}
I_a=a_1\tau^{1/2} \bar{\Delta} + a_2\tau^{3/2}\bar{\nabla}^2\bar{\Delta}+
a_3\tau^{5/2}\bar{\nabla}^2(\bar{\nabla}^2\bar{\Delta}),
\label{eq:Ia_aa}
\end{align}
as appears in Eq.~(\ref{eq:gap_equation_series}).

%%%%%%%%%%%%%%%%%%%%%%%%%%%%%%%%%%%%%%%%%%%%%%%%%%%%%%%%%%%%%%%%%%%%%%%%%%%
\subsection{Coefficients $b_i$ related to the integral kernel $K_b({\bf r},
\{y\}_3)$}
\label{subsec:A2}
%%%%%%%%%%%%%%%%%%%%%%%%%%%%%%%%%%%%%%%%%%%%%%%%%%%%%%%%%%%%%%%%%%%%%%%%%%%

Our next step is to calculate the coefficients $b_i$ in
Eq.~(\ref{eq:gap_equation_series}) that are related to the second
integral kernel, i.e., $K_b({\bf r},\{y\}_3)$. We start with
\begin{multline}
I_b=\frac{1}{g}\int\!\!\prod\limits_{i=1}^3{\rm d}^3z_i\; K_b({\bf
z}_1, {\bf z}_2,{\bf z}_3)\\
\times \Delta({\bf r}+{\bf z}_1) \Delta^{\ast}({\bf r}+{\bf z}_2)
\Delta({\bf r}+{\bf z}_3),
\label{eq:int_Kb}
\end{multline}
with ${\bf z}_i={\bf y}_i-{\bf r}\;(i=\{1,2,3\})$ and $K_b({\bf
z}_1, {\bf z}_2,{\bf z}_3)=K_b({\bf r},\{y\}_3)$. The integral in
Eq.~(\ref{eq:int_Kb}) is expanded in terms of the spatial
derivatives of $\Delta({\bf r})$ and $\Delta^{\ast}({\bf r})$ by
using Eq.~(\ref{eq:gradient_expansion}). The terms that contribute
to the relevant orders $\tau^{3/2}$ and $\tau^{5/2}$ are the
following:
\begin{subequations}\label{eq:Ib}
\begin{align}
&I^{(1)}_b=\frac{\tau^{3/2}}{g}\,\bar{\Delta}\;|\bar{\Delta}|^2\int\!\!
\prod\limits_{i=1}^3{\rm d}^3z_i\; K_b({\bf z}_1,{\bf z}_2,{\bf z}_3),
\label{eq:Ib_1}\\
&I^{(2)}_b=\frac{\tau^{5/2}}{g}\bar{\Delta}\int\!\!\prod\limits_{i=1}^3
{\rm d}^3z_i\;K_b({\bf z}_1,{\bf z}_2,{\bf z}_3)\nonumber\\
&\quad\quad\quad\quad\quad\quad\times({\bf z}_2\cdot\bar{\boldsymbol\nabla})
\bar{\Delta}^{\ast}\;\big(({\bf z}_1 + {\bf z}_3)\cdot\bar{\boldsymbol\nabla}\big)
\bar{\Delta}, \label{eq:Ib_2}\\
&I^{(3)}_b=\frac{\tau^{5/2}}{g}\bar{\Delta}^{\ast}\int\!\!
\prod\limits_{i=1}^3{\rm d}^3z_i\; K_b({\bf z}_1,{\bf z}_2,{\bf z}_3)
\nonumber\\
&\quad\quad\quad\quad\quad\quad\quad\quad\quad\times
({\bf z}_1\cdot\bar{\boldsymbol\nabla})\bar{\Delta}\;\;({\bf z}_3\cdot\bar{\boldsymbol\nabla})\bar{\Delta},
\label{eq:Ib_3}\\
&I^{(4)}_b=\frac{\tau^{5/2}}{g}\,\frac{\bar{\Delta}^2}{2}\int\!\!
\prod\limits_{i
=1}^3{\rm d}^3z_i\; K_b({\bf z}_1,{\bf z}_2,{\bf z}_3)\nonumber\\
&\quad\quad\quad\quad\quad\quad\quad\quad\quad\quad\quad\quad\quad
\times({\bf z}_2\cdot\bar{\boldsymbol\nabla})^2\bar{\Delta}^{\ast},
\label{eq:Ib_4}\\
&I^{(5)}_b=\frac{\tau^{5/2}}{g}\,\frac{|\bar{\Delta}|^2}{2}
\int\!\!\prod\limits_{i=1}^3{\rm d}^3z_i\; K_b({\bf z}_1,{\bf z}_2,
{\bf z}_3)\nonumber\\
&\quad\quad\quad\quad\quad\quad\quad\;\times\big(({\bf z}_1
\cdot\bar{\boldsymbol\nabla})^2+({\bf z}_3\cdot\bar{\boldsymbol\nabla})^2)\bar{\Delta},
\label{eq:Ib_5}
\end{align}
\end{subequations}
The contribution given by Eq.~(\ref{eq:Ib_1}) appears already in the
standard GL domain, and, so, explanations on evaluating $I^{(1)}_b$
can be found in textbooks, see, e.g., Refs.~\onlinecite{degen} and
\onlinecite{fett}. This term reads
\begin{align}
I^{(1)}_b=-b_1\tau^{3/2}|\bar{\Delta}|^2\bar{\Delta}.
\label{eq:Ib_1a}
\end{align}
The other terms in Eq.~(\ref{eq:Ib}) require a more involved calculational procedure. Below the details of such a procedure are given for $I^{(2)}_b$. As to calculations of $I^{(3)}_b,\,I^{(4)}_b$ and $I^{(5)}_b$, we restrict ourselves to only basic remarks.

The term $I^{(2)}_b$ can be written as
\begin{align}
I^{(2)}_b&=-\tau^{5/2}T_c\big(1+{\cal O}(\tau)\big) \sum\limits_{nm} \bar{\Delta} \,\;\bar{\nabla}_n\bar{\Delta}\,\;\bar{\nabla}_m\bar{\Delta}^{\ast}\nonumber\\
&\times\sum\limits_{\omega}\int\prod\limits_{j=1}^3{\rm d}^3z_j\;
{\cal G}^{(0)}_{\omega}(-{\bf z}_1)\;{\widetilde {\cal G}}^{(0)}_{
\omega}({\bf z}_1-{\bf z}_2)
\nonumber\\
&\times {\cal G}^{(0)}_{\omega}({\bf z}_2-{\bf z}_3)\,{\widetilde
{\cal G}}^{(0)}_{\omega}({\bf z}_3)\;z_{2m}\,(z_{1n} + z_{3n}),
\label{eq:Ib_2a}
\end{align}
with ${\cal G}^{(0)}_{\omega}({\bf r},{\bf r}')={\cal
G}^{(0)}_{\omega} ({\bf r}-{\bf r}')$ and ${\bf
z}_j=\{z_{j1},z_{j2},z_{j3}\}$. The integral in Eq.~(\ref{eq:Ib_2a})
is reduced by invoking the Fourier transform and applying the
well-known convolution theorem provided that we rearrange the
polynomial in the relevant integrand as
\begin{align}
z_{2m}\,(z_{1n} + z_{3n})& =(-z_{1m})(-z_{1n}) - (-z_{1m}) z_{3n}
 \nonumber\\+&(z_{1m}-z_{2m}) (-z_{1n})-
(z_{1m}-z_{2m}) z_{3n}.
\nonumber
\end{align}
Then, one finds
\begin{align}
&I^{(2)}_b=\tau^{5/2}T_c\big(1+{\cal O}(\tau)\big)\nonumber\\
\times&\sum\limits_{nm}\bar{\Delta}\;\,\bar{\nabla}_n
\bar{\Delta}\,\;\bar{\nabla}_m\bar{\Delta}^{\ast}
\sum\limits_{\omega}\int\frac{{\rm d}^3k}{(2\pi)^3}\nonumber\\
\times&\Bigg\{\Big(\partial_m\partial_n\frac{1}{i\hbar
\omega-\xi_k}\Big)\frac{1}{(i\hbar\omega+\xi_k)^2(i\hbar\omega-
\xi_k)}
\nonumber\\
&-\Big(\partial_m\frac{1}{i\hbar\omega-\xi_k}\Big)
\Big(\partial_n\frac{1}{i\hbar\omega+\xi_k}\Big)
\frac{1}{\hbar^2\omega^2+\xi^2_k}
\nonumber\\
&+\Big(\partial_m\frac{1}{i\hbar\omega+\xi_k}\Big)
\Big(\partial_n\frac{1}{i\hbar\omega-\xi_k}\Big)
\frac{1}{\hbar^2\omega^2+\xi^2_k}
\nonumber\\
&-\Big(\partial_m\frac{1}{i\hbar\omega+\xi_k}\Big)
\Big(\partial_n\frac{1}{i\hbar\omega+\xi_k}\Big)
\frac{1}{(i\hbar\omega-\xi_k)^2}\Bigg\},
\label{eq:Ib_2b}
\end{align}
with ${\bf \partial}_{{\bf k}}=\{\partial_1,\partial_2, \partial_3\}$. After straightforward but tedious calculations we further obtain
\begin{align}
I^{(2)}_b&=-\tau^{5/2}T_c\big(1+{\cal O}(\tau)\big) \sum\limits_n
\bar{\Delta}\, \;\bar{\nabla}_n\bar{\Delta}\,\;\bar{\nabla}_n\bar{\Delta}^{\ast}\nonumber\\
\times&\sum\limits_{\omega}\int\!\!\!\frac{{\rm d}^3k}{(2\pi
)^3}\;\frac{\hbar^4k^2_n}{m^2}\,\Big[\frac{2}{(i\hbar\omega-
\xi_k)^4(i\hbar\omega+\xi_k)^2}
\nonumber\\
&\quad\quad\quad\quad\quad\quad\quad+\frac{2}{(i\hbar\omega-\xi_k)^3
(i\hbar\omega+\xi_k)^3}\Big],
\label{eq:Ib_2c}
\end{align}
with ${\bf k}=\{k_1,k_2,k_3\}$. Due to the spherical symmetry of the
term in the parenthesis, the integral in Eq.~(\ref{eq:Ib_2c}) does
not depend on $n$ so that $k^2_n$ can be replaced by $k^2/3$. Then,
making use of the standard approximation ($\xi=\xi_k$)
$$
\int\frac{{\rm d}^3k}{(2\pi)^3}\approx N(0)\int\limits_{-\infty}^{
+\infty}\!\!{\rm d}\xi,
$$
one gets
\begin{align}
I^{(2)}_b=&-\tau^{5/2}T_c\big(1+{\cal O}(\tau)\big)\;\bar{\Delta}
\,|\bar{\boldsymbol\nabla}\bar{\Delta}|^2\;\frac{4\hbar^2}{3m}\,N(0)\nonumber\\
&\times \sum\limits_{\omega}\int\limits_{-\infty}^{+\infty}\!\!
{\rm d}\xi\;\frac{(\xi+\mu)(2\xi^2 + 2i\hbar\omega\xi)}{(\hbar^2\omega^2+\xi^2)^4}.
\label{eq:Ib_2d}
\end{align}
The terms in the numerator of the integrand proportional to an odd
power of $\xi$ do not contribute. The same is related to the terms
proportional to $\omega$ due to the summation over the positive and
negative Matsubara frequencies. So, Eq.~(\ref{eq:Ib_2d}) is further
reduced to
\begin{align}
I^{(2)}_b=&-\tau^{5/2}T_c\big(1+{\cal O}(\tau)\big)\bar{\Delta}
\,|\bar{\boldsymbol\nabla}\bar{\Delta}|^2 \nonumber\\
&\times N(0)\mu\,\frac{4\hbar^2}{3m}\sum\limits_{\omega}
\frac{1}{|\hbar\omega|^5}\int\limits_{-\infty}^{+\infty}\!
\!{\rm d}\alpha\;\frac{2\alpha^2}{(1+\alpha^2)^4},
\label{eq:Ib_2e}
\end{align}
where, we recall, $\hbar\omega=\pi T (2n+1)$. Now, using
$$\sum\limits_{n=0}^{\infty}\frac{1}{(n+\frac{1}{2})^5}=31\zeta(5),
\;\;\int\limits_{-\infty}^{+\infty}\!
\!{\rm d}\alpha\;\frac{2\alpha^2}{(1+\alpha^2)^4}=
\frac{\pi}{8},$$
with $\zeta(\ldots)$ the Riemann zeta-function, we arrive at
\begin{align}
I^{(2)}_b=-2b_2\tau^{5/2}\,\bar{\Delta}
\,|\bar{\boldsymbol\nabla}\bar{\Delta}|^2,
\label{eq:Ib_2f}
\end{align}
with $b_2$ given by Eq.~(\ref{eq:coefficients}).

Based on the calculations of $I^{(2)}_b$, we can further proceed with $I^{(3)}_b, \;I^{(4)}_b$ and $I^{(5)}_b$. The contribution given by $I^{(3)}_b$ can be reduced to
\begin{align}
I^{(3)}_b=&-\tau^{5/2}T_c\big(1+{\cal O}(\tau)\big)\bar{\Delta}^{\ast}
\,(\bar{\boldsymbol\nabla}\bar{\Delta})^2 \nonumber\\
&\times N(0)\mu\,\frac{2\hbar^2}{3m}\sum\limits_{\omega}
\frac{1}{|\hbar\omega|^5}\int\limits_{-\infty}^{+\infty}\!
\!{\rm d}\alpha\;\frac{1}{(1+\alpha^2)^3},
\label{eq:Ib_3a}
\end{align}
with
$$
\int\limits_{-\infty}^{+\infty}\!
\!{\rm d}\alpha\;\frac{1}{(1+\alpha^2)^3}=\frac{3\pi}{8}.
$$
When making the summation over $\omega$, Eq.~(\ref{eq:Ib_2a}) becomes of the form
\begin{align}
I^{(3)}_b=-3b_2\tau^{5/2}\,\bar{\Delta}^{\ast}
\,(\bar{\boldsymbol\nabla}\bar{\Delta})^2.
\label{eq:Ib_3b}
\end{align}

For $I^{(4)}_b$ we obtain
\begin{align}
I^{(4)}_b=&\tau^{5/2}T_c\big(1+{\cal O}(\tau)\big)\bar{\Delta}^2
\,\bar{\nabla}^2\bar{\Delta}^{\ast} \nonumber\\
&\times N(0)\mu\,\frac{2\hbar^2}{3m}\sum\limits_{\omega}
\frac{1}{|\hbar\omega|^5}\int\limits_{-\infty}^{+\infty}\!
\!{\rm d}\alpha\;\frac{3\alpha^2-1}{(1+\alpha^2)^4},
\label{eq:Ib_4a}
\end{align}
with
$$
\int\limits_{-\infty}^{+\infty}\!
\!{\rm d}\alpha\;\frac{3\alpha^2-1}{(1+\alpha^2)^3}=-\frac{\pi}{8}.
$$
This results in
\begin{align}
I^{(4)}_b=-b_2\tau^{5/2}\,\bar{\Delta}^2
\,\bar{\bf \nabla}^2\bar{\Delta}^{\ast}.
\label{eq:Ib_4b}
\end{align}

At last, $I^{(5)}_b$ is reduced to
\begin{align}
I^{(5)}_b=&\tau^{5/2}T_c\big(1+{\cal O}(\tau)\big)|\bar{\Delta}|^2
\,\bar{{\bf \nabla}}^2\bar{\Delta} \nonumber\\
&\times N(0)\mu\,\frac{4\hbar^2}{3m}\sum\limits_{\omega}
\frac{1}{|\hbar\omega|^5}\int\limits_{-\infty}^{+\infty}\!
\!{\rm d}\alpha\;\frac{\alpha^2-1}{(1+\alpha^2)^4},
\label{eq:Ib_5a}
\end{align}
which results in
\begin{align}
I^{(5)}_b=-4b_2\tau^{5/2}\,|\bar{\Delta}|^2
\,\bar{\nabla}^2\bar{\Delta}^{\ast}.
\label{eq:Ib_5b}
\end{align}

Now, based on Eqs.~(\ref{eq:int_Kb}), (\ref{eq:Ib_1a}), (\ref{eq:Ib_2f}),
(\ref{eq:Ib_3b}), (\ref{eq:Ib_4b}) and (\ref{eq:Ib_5b}), we obtain
\begin{align}
I_b=&-b_1\tau^{3/2}|\bar{\Delta}|^2\bar{\Delta}
- b_2\tau^{5/2} \Big[2\bar{\Delta}\;|\bar{\boldsymbol\nabla}
\bar{\Delta}|^2\nonumber \\
&+3\bar{\Delta}^\ast (\bar{\boldsymbol\nabla} \bar{\Delta})^2
+\bar{\Delta}^2\; \bar{\bf \nabla}^2\bar{\Delta}^{\ast}
+4|\bar \Delta|^2 \bar{\bf \nabla}^2\bar{\Delta}\Big],
\label{eq:Iba}
\end{align}
with $b_1$ and $b_2$ defined by Eq.~(\ref{eq:coefficients}).

%%%%%%%%%%%%%%%%%%%%%%%%%%%%%%%%%%%%%%%%%%%%%%%%%%%%%%%%%%%%%%%%%%%%%%%%%%
\subsection{Coefficient $c_1$ coming from $K_c({\bf r},\{y\}_5)$}
\label{subsec:A3}
%%%%%%%%%%%%%%%%%%%%%%%%%%%%%%%%%%%%%%%%%%%%%%%%%%%%%%%%%%%%%%%%%%%%%%%%%%

The term with the coefficients $c_1$ in Eq.~(\ref{eq:gap_equation_series})
appears due to the contribution to $\Delta({\bf r})/g$ given by
\begin{align}
I_c&=\int\!\!\prod_{j=1}^5 {\rm d}^3z_j \;K_c({\bf z}_1,{\bf z}_2,
{\bf z}_3,
{\bf z}_4,{\bf z}_5)\;\Delta({\bf r} +{\bf z}_1)\notag \\
& \times \Delta^{\ast}({\bf r} +{\bf z}_2)\Delta({\bf r} +{\bf z}_3)
\Delta^{\ast}({\bf r} +{\bf z}_4)\Delta({\bf r} +{\bf z}_5).
\label{eq:Ic}
\end{align}
We need all contributions up to the order $\tau^{5/2}$ in
Eq.~(\ref{eq:gap_equation_series}). As the leading-order term in the
order parameter is proportional to $\tau^{1/2}$, it is possible to
neglect the contribution of the spatial derivatives of the order
parameter and limit ourselves only to the local contribution given
by
\begin{align}
I^{(1)}_c=\tau^{5/2}&\big(1+{\cal O}(\tau)\big)\;\bar{\Delta}\,
|\bar{\Delta}|^4\nonumber\\
&\times\int\!\!\prod_{j=1}^5 {\rm d}^3z_j \;K_c({\bf z}_1,{\bf z}_2,
{\bf z}_3,{\bf z}_4,{\bf z}_5).
\label{eq:Ic_1}
\end{align}
Using Eq.~(\ref{eq:kernels}), performing the Fourier transformation,
and passing to  the integration over the single-electron energy, we
can find
\begin{align}
I^{(1)}_c&=\tau^{5/2}T_c\big(1+{\cal O}(\tau)\big)\;\bar{\Delta}
\,|\bar{\Delta}|^4\; N(0)\nonumber\\
&\quad\quad\quad\times\sum\limits_{\omega}\frac{1}{|\hbar\omega|^5}
\int\limits_{-\infty}^{+\infty}\!\!
{\rm d}\alpha\;\frac{1}{(\alpha^2+1)^3},
\label{eq:Ic_1a}
\end{align}
where the integral is equal to $3\pi/8$, see the previous
subsection. Evaluating the sum over the Matsubara frequencies in
Eq.~(\ref{eq:Ic_1a}), we obtain
\begin{align}
I_c&=c_1\tau^{5/2}\;\bar{\Delta}\,|\bar{\Delta}|^4,
\label{eq:Ica}
\end{align}
where $c_1$ is given by Eq.~(\ref{eq:coefficients}).

%%%%%%%%%%%%%%%%%%%%%%%%%%%%%%%%%%%%%%%%%%%%%%%%%%%%%%%%%%%%%%%%%%%%%%%%%%%%%
\section{The normal-state Green function in the presence of a magnetic field}
\label{sec:appendix_B}
%%%%%%%%%%%%%%%%%%%%%%%%%%%%%%%%%%%%%%%%%%%%%%%%%%%%%%%%%%%%%%%%%%%%%%%%%%%%%

As discussed in the main text of the article, constructing the EGL formalism requires the calculation of the normal-state Green function in the presence of a magnetic field with accuracy ${\cal O}(\tau^2)$. This means that we are not able to rely upon the phase-integral approximation of Gor'kov given by
Eq.~(\ref{eq:green_B}), where the classical particle trajectory is assumed to be a straight line. To go beyond this approximation, we need to take $\tau$-dependent deviations from such a linear trajectory. A natural way of doing so is based on the following representation for the single-particle propagator (see, e.g.,
Ref.~\onlinecite{zagoskin}):
\begin{align}
{\cal G}^{(0)}({\bf r}t, {\bf r}'t') = F(t-t')\;e^{\frac{i}{\hbar}
S_{\rm cl}},
\label{eq_Gf}
\end{align}
where $S_{\rm cl}$ is the classical action
\begin{equation}
S_{\rm cl} =\int\limits_{t'}^t\!\frac{m {\dot {\bf q}}^2}{2} ds +
\frac{e}{\mathfrak{c}}\int\limits_C\!{\bf A}({\bf q})\cdot {\rm d}{\bf q},
\label{eq_Scl}
\end{equation}
where the integrals are taken along the classical trajectory $C$
that satisfies the equation of motion
\begin{equation}
\ddot{\bf q} = \frac{e}{m\,\mathfrak{c}} (\dot{\bf q} \times {\bf B}),
\label{eq_motion}
\end{equation}
with the boundary conditions ${\bf q}(t')={\bf r}',\,{\bf q}(t)=
{\bf r}$. We are interested in the systematic corrections to Gor'kovs
eikonal approximation and, so, it is of convenience to recast
$S_{\rm cl}$ in the form:
\begin{equation}
S_{\rm cl} = S_{\rm Gor} + \int\limits_{t'}^t\!\!\frac{m}{2}\Big[
{\dot{\bf q}}^2 -\Big(\frac{{\bf r}-{\bf r}'}{t-t'}\Big)^2\Big] {\rm
d}s  + \frac{e}{\mathfrak{c}} \iint\limits_{\Sigma} {\bf B}({\bf q})
\cdot{\rm d}{\bf\Sigma}, \label{eq_Scl1}
\end{equation}
where $S_{\rm Gor}$ is the classical action along the straight line
connecting ${\bf r}'$ and ${\bf r}$, i.e.,
\begin{equation}
S_{\rm Gor}=\frac{m ({\bf r}-{\bf r}')^2}{2(t-t')} +
\frac{e}{\mathfrak{c}}\int\limits_{{\bf r}'}^{\bf r}\! {\bf A}({\bf
q})\cdot{\rm d}{\bf q}, \label{eq_SGor}
\end{equation}
which is the basis of the Gor'kovs approximation for the normal-state
Green function in a magnetic field. The integral in
Eq.~(\ref{eq_Scl}) is over the surface $\Sigma$ that is bound by the
loop $\partial\Sigma$ consisting of two parts, i.e., the classical
trajectory $C$ from ${\bf r}'$ to ${\bf r}$ and the straight line
connecting ${\bf r}$ and ${\bf r}'$. The orientation of the $\partial
\Sigma$ is positive with respect to the surface.

To simplify the further analysis, we introduce the decomposition ${\bf
q}={\bf q}_1 + {\bf q}_2$ and recast Eq.~(\ref{eq_motion}) as follows:
\begin{subequations}
\begin{align}
&\ddot{\bf q}_1 =\frac{e}{m\,\mathfrak{c}}\,\big(\dot{\bf q}_1\times
{\bf B}({\bf r})\big),
\label{eq_motion_A}\\
&\ddot{\bf q}_2 =\frac{e}{m\,\mathfrak{c}}\,\big(\dot{\bf q}_1\times
\big[{\bf B}({\bf q})-{\bf B}({\bf r})\big]\big) + \frac{e}{m\,
\mathfrak{c}}\,\big(\dot{\bf q}_2 \times {\bf B}
({\bf q})\big),
\label{eq_motion_B}
\end{align}
\end{subequations}
with the boundary conditions ${\bf q}_1(t')={\bf r}'$, ${\bf q}_1(t)
={\bf r}$ and ${\bf q}_2(t)={\bf q}_2(t')=0$. This decomposition is
such that ${\bf q}_1$ is the solution for the uniform (not dependent
on ${\bf q}$) magnetic field ${\bf B}({\bf r})$. This rearrangement
is convenient because, as seen below, the spatial derivatives of
${\bf B}({\bf r})$ contribute to the propagator only in the order
$\tau^{5/2}$. Now, we set ${\bf B}({\bf r})= B({\bf r}){\bf e}_z$,
with ${\bf e}_z$ the unit vector in the $z$-direction. The corresponding $\tau$-expansion for ${\bf q}_1$ can be found from straightforward
calculations with the result given by (for the moment, we put ${\bf r}'=0,
t'=0$)
\begin{subequations}\label{eq_motion_A1}
\begin{align}
&q_{1x}= x\,s/t + \tau\, y\bar\Omega\varphi(s)+ \tau^2\,x\bar\Omega^2
\chi(s) + {\cal O}(\tau^3),\label{eq_motion_A1a}\\
&q_{1y}= y\,s/t - \tau\,x\bar\Omega\varphi(s)+ \tau^2\,y\bar\Omega^2
\chi(s)+
{\cal O}(\tau^3),
\label{eq_motion_A1b}\\
&q_{1z}= z\,s/t,
\label{eq_motion_A1c}
\end{align}
\end{subequations}
where $\bar\Omega=|e|\bar B({\bf r})/m\mathfrak{c}$, with the scaled
magnetic field $\bar{B} ({\bf r}) = \frac{1}{\tau} B({\bf r})$~(in
other words, $\bar{\Omega}$ stands for the scaled cyclotron
frequency). In addition, ${\bf q}_1=\{q_{1x},q_{1y},q_{1z}\}$ and
${\bf r}=\{x,y,z\}$ in Eq.~(\ref{eq_motion_A1}), and $\varphi (s)$
and $\chi(s)$ are given by
$$
\varphi(s)=\frac{s}{2}\Big(1-\frac{s}{t}\Big),\;\chi(s)=
\frac{s^2}{4} \Big(1 - \frac{t}{3s} - \frac{2s}{3t}\Big),
$$
where, as seen, $\varphi(t)=\chi(t)=0$. It is worth noting that
$\bar B({\bf r})$ does not depend on  $\tau$, see
Eq.~(\ref{eq:field_expansion}).

The solution of Eq.~(\ref{eq_motion_B}) is more complicated because
it involves the spatial derivatives of ${\bf B}({\bf r})$. Let
us first consider a simplified variant when the magnetic field is
generally parallel to the $z$-axis while its absolute value varies
with position, i.e., ${\bf B}({\bf q})= B({\bf q}){\bf e}_z$. Then,
Eq.~(\ref{eq_motion_B}) can be rearranged as
\begin{equation}
\ddot{\bf q}_2 =\frac{e}{m\,\mathfrak{c}}\,\big[B({\bf q}_1^{(0)})-
B({\bf r})\big](\dot{\bf q}^{(0)}_1\times {\bf e}_z\big)+{\cal O}
(\tau^{5/2}),
\label{eq_motion_B1}
\end{equation}
with ${\bf q}^{(0)}_1 =\lim_{\tau \to 0} {\bf q}_1$, i.e., ${\bf
q}^{(0)}_1 =\{xs/t,ys/t,zs/t\}$. We can further simplify this equation by making use of
\begin{align}
&B({\bf q}_1^{(0)})-B({\bf r}) =\nonumber
\\&=\tau^{3/2}\,\frac{s}{t}({\bf r}\cdot\bar{\boldsymbol\nabla}){\bar B}({\bf r})
+ \tau^2\frac{s^2}{2\,t^2}({\bf r}\cdot\bar{\boldsymbol\nabla})^2\bar{B}({\bf r}) +
{\cal O}(\tau^{5/2}), \nonumber
\end{align}
where scaled derivatives are introduced, i.e., $\bar{\bf\nabla}=\tau^{-1/2} \bf{\nabla}$, and the operator $\bar\nabla$ acts only on the magnetic field. From the resulting equation, we obtain
\begin{subequations}\label{eq_motion_B2}
\begin{align}
&q_{2x}=\frac{y}{t}\;\Xi(s,{\bf r}) + {\cal O}(\tau^{5/2}),
\label{eq_motion_B2a}\\
&q_{2y}=-\frac{x}{t}\;\Xi(s,{\bf r}) + {\cal O}(\tau^{5/2}),
\label{eq_motion_B2b}\\
&q_{2z}=0 \label{eq_motion_B2c},
\end{align}
\end{subequations}
where $\Xi(s,{\bf r})$ satisfies the differential equation
$$
\ddot\Xi(s,{\bf r})=\frac{e}{m\,\mathfrak{c}} \big[\tau^{3/2}\,
\frac{s}{t}({\bf r}\cdot\bar{\boldsymbol\nabla}){\bar B}({\bf r}) + \tau^2
\frac{s^2}{2t^2}({\bf r}\cdot\bar{\boldsymbol\nabla})^2\bar{B}({\bf r})\big],
$$
supplemented by the boundary conditions $\Xi(0,{\bf r})=
\Xi(t,{\bf r})=0$.

Now, we have everything at our disposal to calculate $S_{\rm cl}-
S_{\rm Gor}$ and, so, to analyze the corrections to the approximation
employed by Gor'kov. Using Eqs.~(\ref{eq_Scl}), (\ref{eq_SGor}), (\ref{eq_motion_A1}) and (\ref{eq_motion_B2}), we find
\begin{equation}
S_{\rm cl} - S_{\rm Gor}= -\frac{m}{24}\tau^2\,(x^2+y^2)\bar{\Omega}^2\,t
+ {\cal O}(\tau^{5/2}). \label{eq_Scl2}
\end{equation}
It is remarkable that the spatial derivatives of the magnetic field
contribute to $S_{\rm cl}-S_{\rm Gor}$ only in orders higher than
$\tau^2$. In particular, when considering the surface
integral in Eq.~(\ref{eq_Scl1}), ${\bf B} \propto \tau$ and the
contribution of the spatial derivatives of the  magnetic field to
the surface integral, taken in its lowest order, is proportional to
$\tau^{3/2}$. So, the resulting product is of the order $\tau^{5/2}$,
which means that it does not make a contribution to the first term
in the right-hand-side of Eq.~(\ref{eq_Scl2}). In turn, calculating
the kinetic energy we find
$$
{\dot{\bf q}}^2 = {\dot{\bf q}_1}^2 + 2\,\frac{\Xi(s,{\bf r})}{t}
\Big(y\dot{q}_{1x}-x\dot{q}_{1y}\Big) + {\cal O}(\tau^{5/2}),
$$
which, taken together with $\dot{q}_{1x}=x/t$ and $\dot{q}_{1y}=y/t$,
makes it possible to conclude that the spatial derivatives of the
magnetic field can contribute to $S_{\rm cl}$ only to the order
$\tau^{5/2}$.

Let us make a few remarks about Eq.~(\ref{eq_Scl2}). Though this
result was derived under the simplified assumption ${\bf B}({\bf
q})=B({\bf q}){\bf e}_z$, it is general and holds even in the
presence of spatial variations in the direction of the magnetic field.
Thiscan be seen from the following. Based on our above consideration,
we expect that the two lowest orders contributing to ${\bf B}({\bf
q}^{(0)}_1)-{\bf B}({\bf r})$ are $\tau^{3/2}$ and $\tau^2$. Then,
Eq.~(\ref{eq_motion_B}) can be reduced to
\begin{equation}
\ddot{\bf q}_2 =\frac{e}{m\,\mathfrak{c}}\,\big(\dot{\bf q}^{(0)}_1
\times\big[{\bf B}({\bf q}^{(0)}_1)-{\bf B}({\bf r})\big]\big) +
{\cal O}(\tau^{5/2}),
\label{eq_motion_B3}
\end{equation}
whose solution reads ($i=\{x,y,z\}$)
\begin{equation}
q_{2,i}=\frac{e}{m\,\mathfrak{c}\,t} \sum\limits_{jk} \varepsilon_{ijk}
\,r_j\Upsilon_k(s,{\bf r}) + {\cal O}(\tau^{5/2}),
\label{eq_motion_B4}
\end{equation}
with $\varepsilon_{jmk}$ the permutation tensor and ${\bf B}_i({\bf
q}^{(0)}_1)-{\bf B}_i({\bf r})=\ddot{\Upsilon}_i$, where
$\Upsilon_i$ is taken in the two lowest orders in $\tau$~(with the
boundary conditions $\Upsilon_i(0,{\bf r})=\Upsilon_i(t,{\bf
r})=0$). Whatever $\Upsilon_i$, it does not make a contribution
neither to the surface integral nor to the kinetic term in each
order lower than $\tau^{5/2}$. In particular, in the kinetic term we
obtain
$$
{\dot{\bf q}}^2 ={\dot{\bf q}_1}^2 + \frac{2}{t}\sum\limits_{ijk}
\varepsilon_{ijk}r_i r_k \Upsilon_j(s,{\bf r}) + {\cal
O}(\tau^{5/2}),
$$
where the second term is simply equal to zero. We also remark that
detailed calculations make it possible to find that
\begin{equation}
{\bf q}_2=\frac{e\tau^{3/2}}{6m\,\mathfrak{c}}\, s\Big(1-
\frac{s^2}{t^2}\Big)({\bf r}\cdot \bar{\boldsymbol\nabla})\big(\bar{\bf B}
({\bf r})\times{\bf r}\big) + {\cal O}(\tau^2),
\label{eq_motion_B5}
\end{equation}
which means that the above assumption about the two lowest contributing
orders in ${\bf B}({\bf q}^{(0)}_1)-{\bf B}({\bf r})$ is fully correct.

The only thing remaining is to specify the quantum-fluctuation factor $F(t)$~($t'=0$ is still of use). In the Gor'kov approximation $F(t)= \left(\frac{m}{2 \pi i \hbar t} \right)^{3/2}$ but this is not longer
the case in the order $\tau^2$ and higher. Within accuracy ${\cal O}
(\tau^2)$ we obtain
\begin{align}
F(t) = \Big(\frac{m}{2\pi i \hbar t} \Big)^{3/2} \left( 1 +
\frac{\tau^2{\bar\Omega}^2}{24}\,t^2\right), \label{eq_fluct_factor}
\end{align}
which is nothing else but the fluctuation factor for the propagator
in a uniform magnetic field expanded in $\tau$ (see, e.g., Ref.~\onlinecite{prop}).

Now, based on Eqs.~(\ref{eq_Scl1}), (\ref{eq_SGor}), (\ref{eq_Scl2}) and (\ref{eq_fluct_factor}) and making the usual imaginary-time substitution
$t \to -i t$, we arrive at the following expression for the temperature single-particle Green function in the presence of a magnetic field ($t
\to t-t'$ and ${\bf r} \to {\bf r}-{\bf r}'$):
\begin{align}
&{\cal G}^{(0)}({\bf r}t, {\bf r}'t') ={\cal G}^{(0)}_{\rm Gor}({\bf
r}t,{\bf r}'t')\left\{1 - \frac{\tau^2 \bar{\Omega}^2}{24}\Big[(t-t')^2
\right.\nonumber\\
&\left.\quad\quad\quad\quad+\frac{m}{\hbar}\big({\bf r}-{\bf r}'
\big)_{\perp}^2\,(t-t')\Big] + {\cal O}(\tau^{5/2})\right\},
\label{eq_Gf1}
\end{align}
where $({\bf r}-{\bf r}')_{\perp}$ is the component of the vector ${\bf r}-{\bf r}'$ perpendicular to ${\bf B}({\bf r})$ and ${\cal G}^{(0)}_{\rm Gor}({\bf r}t,{\bf r}'t')$ is the Gor'kov approximation for the normal-state Green function given by Eq.~(\ref{eq:green_B}). Now, switching to the Fourier transform of
the Green function given by Eq.~(\ref{eq_Gf1}) we find Eq.~(\ref{eq:Green_B_corr}). As seen from Eqs.~(\ref{eq:Green_B_corr}) and (\ref{eq_Gf1}), the spatial
derivatives contribute to the $\tau$-expansion of the temperature normal-state Green function only in the order $\tau^{5/2}$ and higher. However, to construct the EGL formalism (up to the order $\tau^{3/2}$ in the order parameter), we need to know ${\cal G}^{(0)}_{\omega}({\bf r},{\bf r}')$ only up to the order $\tau^2$.

%%%%%%%%%%%%%%%%%%%%%%%%%%%%%%%%%%%%%%%%%%%%%%%%%%%%%%%%%%%%%%%%%%%%%%%%%%%%%%%%%%
\section{Expansion for the gap equation in the presence of a nonzero
magnetic field}
\label{sec:appendix_C}
%%%%%%%%%%%%%%%%%%%%%%%%%%%%%%%%%%%%%%%%%%%%%%%%%%%%%%%%%%%%%%%%%%%%%%%%%%%%%%%%%%

In Appendix \ref{sec:appendix_A} we gave the details of the calculations for the coefficients appearing in the $\tau$-expansion of the gap equation. It is of great importance to specify complications that appear when generalizing the procedure to the case of a nonzero magnetic field, i.e., for the normal-state temperature Green function given by Eq.~(\ref{eq:Green_B_corr}).

%%%%%%%%%%%%%%%%%%%%%%%%%%%%%%%%%%%%%%%%%%%%%%%%%%%%%%%%%%%%%%%%%%%%%%%%%%%%%%%%%%
\subsection{Terms related to the integral kernel $K_a({\bf r},{\bf y})$}
\label{subsec:C1}
%%%%%%%%%%%%%%%%%%%%%%%%%%%%%%%%%%%%%%%%%%%%%%%%%%%%%%%%%%%%%%%%%%%%%%%%%%%%%%%%%%

When switching to a nonzero magnetic field, Eq.~(\ref{eq:int_Ka}) in Appendix \ref{sec:appendix_A} can be rewritten in the form
\begin{align}
I_a = &\frac{1}{g}\int\!\! {\rm d}^3y K_a({\bf r},{\bf y})\Delta({\bf y})\nonumber\\
=&\lim\limits_{{\bf r}'\to {\bf r}}\frac{1}{g}\int\!\!{\rm d}^3z\,
Q_a({\bf r},{\bf z})\Delta({\bf r}+{\bf z},{\bf r}'),
\label{eq:int_Ka_B}
\end{align}
where the ``two-point" order parameter $\Delta({\bf r},{\bf r}')$ is
defined by Eq.~(\ref{eq:order_parameter_modified}) and
\begin{align}
&Q_a({\bf r},{\bf z})=K_{a,B=0}({\bf z})-gT_c\,\frac{e^2{\bf
B}^2({\bf r})}{24m^2
\mathfrak{c}^2}\nonumber\\
&\times \sum\limits_{\omega}\Big\{{\cal G}^{(0)}_{\omega,B=0}(-{\bf
z})\big(\partial_{\omega}^2 -\frac{i}{\hbar}m \,{\bf z}^2_{\perp}
\partial_{\omega}\big){\widetilde {\cal
G}}^{(0)}_{\omega, B=0}({\bf z})\nonumber\\
&+{\widetilde {\cal G}}^{(0)}_{\omega,B=0}({\bf
z})\big(\partial_{\omega}^2 + \frac{i}{\hbar} m \,{\bf
z}^2_{\perp}\partial_{\omega}\big){\cal G}^{(0)}_{\omega,B=0}(-{\bf
z})\Big\} + {\cal O}(\tau^{5/2}), \label{eq:Ka_B}
\end{align}
with $K_{a,B=0}({\bf z})$ the zero-magnetic field kernel $K_a$ given
by Eqs.~(\ref{eq:green_0}) and (\ref{eq:kernels}).  Then,
introducing the expansion of $\Delta({\bf y},{\bf r}')$ in powers of
${\bf z}={\bf y}-{\bf r}$ and collecting the relevant orders in
$\tau$, we obtain the following contributions:
\begin{subequations}\label{eq:IaB}
\begin{align}
&I^{(1)}_a=\frac{\tau^{1/2}}{g}\lim\limits_{{\bf r}'\to {\bf
r}}\bar{\Delta}(\bar{\bf r},\bar{\bf r}')\int\!\!{\rm d}^3z\,
K_{a,B=0}({\bf z}),\label{eq:Ia_1B}\\
&I^{(2)}_a=\frac{\tau^{3/2}}{g}\lim\limits_{{\bf r}'\to {\bf
r}}\int\!\!{\rm d}^3z\, K_{a,B=0}({\bf z}) \frac{({\bf z}\cdot\bar{\boldsymbol
\nabla}_{\bf r})^2}{2!}\bar{\Delta}(\bar{\bf r},\bar{\bf r}'),
\label{eq:Ia_2B}\\
&I^{(3)}_a=\frac{\tau^{5/2}}{g}\lim\limits_{{\bf r}'\to {\bf
r}}\int\!\!{\rm d}^3z\, K_{a,B=0}({\bf z}) \frac{({\bf z}\cdot\bar{\boldsymbol
\nabla}_{\bf r})^4}{4!}\bar{\Delta}(\bar{\bf r},\bar{\bf r}'),
\label{eq:Ia_3B}\\
&I^{(4)}_a=-\tau^{5/2}T_c\;\frac{e^2\bar{\bf B}^2(\bar{\bf
r})}{12m^2 \mathfrak{c}^2}\lim\limits_{{\bf r}'\to {\bf
r}}\bar{\Delta}(\bar{\bf
r},\bar{\bf r}')\nonumber\\
&\times\int\!\!{\rm d}^3z\, \sum\limits_{\omega}{\cal
G}^{(0)}_{\omega,B=0}(-{\bf z})\big(\partial_{\omega}^2
-\frac{i}{\hbar}m \,{\bf z}^2_{\perp}
\partial_{\omega}\big){\widetilde {\cal
G}}^{(0)}_{\omega, B=0}({\bf z}).
\end{align}
\end{subequations}
After integrating over ${\bf z}$ (for $I^{(1)}_a, \,I^{(2)}_2,
\,I^{(3)}_a$ see details in Appendix \ref{sec:appendix_A}; for
$I^{(4)}_a$ see the discussion below), Eqs.~(\ref{eq:IaB}) are
reduced to
\begin{subequations}\label{eq:IaB_add}
\begin{align}
&I^{(1)}_a=a_1\tau^{1/2}\lim\limits_{{\bf r}'\to{\bf r}}\bar{\Delta}
(\bar{\bf r},\bar{\bf r}'),\label{eq:Ia_1B_add}\\
&I^{(2)}_a=a_2\tau^{3/2} \lim\limits_{{\bf r}'\to{\bf
r}}\bar{\bf\nabla}_{\bf r}^2 \bar{\Delta}(\bar{\bf r},\bar{\bf r}'),
\label{eq:Ia_2B_add}\\
&I^{(3)}_a=a_3\tau^{5/2}\lim\limits_{{\bf r}'\to{\bf r}}\bar{\bf
\nabla}_{\bf r}^2 \big(\bar{\bf \nabla}_{\bf r}^2
\bar{\Delta}(\bar{\bf r},\bar{\bf r}')\big), \label{eq:Ia_3B_add}\\
&I^{(4)}_a=-a_4\tau^{5/2}\bar{\bf B}^2(\bar{\bf r})\lim\limits_{{\bf
r}'\to {\bf r}}\bar{\Delta}(\bar{ \bf r},\bar{\bf r}'),
\label{eq:Ia_4B_add}
\end{align}
\end{subequations}
where the coefficients $a_1,\,a_2,\,a_3$ are given by
Eq.~(\ref{eq:coefficients}) and $a_4$ is defined in
Eq.~(\ref{eq:a4}).

As already mentioned in Appendix \ref{sec:appendix_A}, the first
two terms, i.e., $I^{(1)}_a$ and $I^{(2)}_a$ appear even in the
standard GL domain and, so, the details of calculating these
contributions are well-known from textbooks.~\cite{degen,fett}
The results are given by Eqs.~(\ref{eq:nabla_shiftedA_a}) and (\ref{eq:nabla_shiftedA_b}), and the only difference from the
standard GL theory is that the coefficients $a_1$ and $a_2$
contains now extra terms of the order $\tau^2$. Calculating
$I^{(3)}_a$ is a more involved and complicated task and, so,
the basic details of calculating $I^{(3)}_a$ are outlined below.
We remark that
\begin{multline}
\lim\limits_{{\bf r}'\to{\bf r}}\big(\bar{\bf \nabla}_{\bf
r}^2\big)^2\bar\Delta(\bar{\bf r},\bar{\bf r}')=
\\=\lim\limits_{{\bf
r}'\to{\bf r}}\Big[\Big(\bar{\boldsymbol\nabla}_{\bf r} -\frac{2ie}{ \hbar\,
\mathfrak{c}}\Phi'_{\bf r}(\bar{\bf r},\bar{\bf
r}')\Big)^2\Big]^2\Delta({\bf r}),
\end{multline}
with $\Phi'_{\bf r}(\bar{\bf r},\bar{\bf r}')=\bar{\boldsymbol\nabla}_{\bf
r}\Phi(\bar{\bf r},\bar{\bf r}')$, where
\begin{align}
\Phi(\bar{\bf r},\bar{\bf r}')=\int\limits_{\bar{\bf r}'}^{\bar{\bf
r}}\bar{ \bf A}\cdot{\rm d}\bar{\bf q}, \label{eq:Phi}
\end{align}
here the integral is taken along the straight line connecting the
points ${\bf r}'$ and ${\bf r}$. Expanding ${\bf A}({\bf q})$ in
powers of $\bar{\bf q}-\bar{\bf r}'$ we can rewrite $\Phi(\bar{\bf r},
\bar{\bf r}')$ in the form
\begin{align}
\Phi(\bar{\bf r},\bar{\bf r}')=\sum\limits_{n=0}^{\infty}
\frac{({\bf a}\cdot\bar{\boldsymbol\nabla}_{\bf r'})^n}{(n+1)!}\big({\bf
a}\cdot\bar{\bf A}(\bar{\bf r}')\big)\Big|_{{\bf a}=\bar{\bf r}-\bar{\bf
r}'}. \label{eq:phase_exp}
\end{align}
As seen, $\lim\limits_{{\bf r}'\to{\bf r}}{\boldsymbol\nabla}_{\bf r}
\Phi(\bar{\bf r},\bar{\bf r}') = \bar{\bf A}(\bar{\bf r})$ and,
so, one could expect that $\lim\limits_{{\bf r}'\to{\bf r}}\big(\bar{\bf \nabla}_{\bf r}^2\big)^2=\big(\bar{\bf D}^2\big)^2$ for $n=1,2$ in Eqs.~(\ref{eq:Ia_2B_add}) and (\ref{eq:Ia_3B_add}). However, this
is true only for $n=1$~(i.e., for $I^{(2)}_a$) and does not hold
for $n=2$: the limiting procedure and differentiating do not
commute in general. In particular, straightforward but tedious
calculations show that
\begin{align}
&\lim\limits_{{\bf r}'\to{\bf r}}\big(\bar{\bf \nabla}_{\bf r}^2\big)^2\Big(
\bar{\Delta}(\bar{\bf r})\; e^{\textstyle -\frac{2ie}{ \hbar\,
\mathfrak{c}}\Phi(\bar{\bf r},\bar{\bf r}')}\Big)\nonumber\\
&=\bar{\bf D}^2\big(\bar{\bf D}^2\bar{\Delta}\big)-\frac{4ie}{3\hbar\,
\mathfrak{c}}\sum\limits_{ij} \bar{\bf \nabla}_i
\bar{\Delta}\big(\bar{\bf \nabla}_i\bar{\bf
\nabla}_j\bar{A}_j-\bar{\bf \nabla}_j^2 \bar{A}_i\big)\nonumber\\
&\;\quad\quad\quad-\frac{4e^2}{\hbar^2\mathfrak{c}^2}\bar{\Delta}
\sum\limits_{ij}\Big(\bar{\bf \nabla}_i\bar{A}_j\big(\bar{\bf
\nabla}_j\bar{A}_i-\bar{\bf \nabla}_i \bar{A}_j\big)\nonumber\\
&\;\quad\quad\quad\quad\quad\quad\quad-\frac{2}{3}\bar{A}_j(\bar{\bf
\nabla}^2_i\bar{A}_j-\bar{\bf \nabla}_j \bar{\bf
\nabla}_i\bar{A}_i)\Big),
\end{align}
with $\bar{\bf A}=\{\bar{A}_1,\bar{A}_2,\bar{A}_3\}$ and
$\bar{\boldsymbol\nabla}=\{\bar{\bf \nabla}_1,\bar{\bf \nabla}_2,\bar{\bf
\nabla}_3 \}$. The above expression results immediately in
Eq.~(\ref{eq:nabla_shifted_c}), when keeping in mind that
$\nabla_iA_j-\nabla_jA_i=\sum_k \varepsilon_{ijk}B_k$, with
$\varepsilon_{ijk}$ the permutation tensor and ${\bf
B}=\{B_1,B_2,B_3\}$.

Concluding this subsection, we note that integration over ${\bf z}$
in $I^{(4)}_a$~(together with the accompanying summation over the
Matsubara frequencies) is similar to that for $I^{(i)}_a$ with
$i=1,2,3$. Invoking the Fourier transformation and converting ${\bf
z}^2_{\perp}$ into the corresponding derivatives of the Green
functions with respect to the single-particle energy, for
$I^{(4)}_a$ we find
\begin{align}
&I^{(4)}_a=-\tau^{5/2}T_c\big(1+{\cal O}(\tau)\big)\,
\frac{e^2\bar{\bf B}^2(\bar{\bf r})}{12m^2\mathfrak{c}^2}
\lim\limits_{{\bf r}'\to {\bf r}}\bar{\Delta}(\bar{ \bf r},\bar{\bf
r}')\,\nonumber\\
&\;\times\sum\limits_{\omega}\int\frac{{\rm d}^3k}{(2\pi)^3}
\Big\{\frac{(-2\hbar^2)}{(i\hbar\omega+\xi_k)(i\hbar\omega-\xi_k)^3}
\nonumber\\
&\quad\quad\quad\quad+\frac{m}{(i\hbar\omega-\xi_k)^2}\;(-\partial^2_1
-\partial^2_2)\frac{1}{i\hbar\omega+\xi_k}\Big\},
\end{align}
with $\partial_{\bf k}=\{\partial_1,\partial_2,\partial_3\}$. This is
reduced to
\begin{align}
&I^{(4)}_a=-\tau^{5/2}T_c\big(1+{\cal O}(\tau)\big)\lim\limits_{{\bf
r}'\to {\bf r}}\bar{\Delta}(\bar{ \bf r},\bar{\bf
r}')\nonumber\\
&\quad\;\times\frac{\hbar^2e^2\bar{\bf B}^2(\bar{\bf r})}{9m^2
\mathfrak{c}^2}N(0) \sum\limits_{\omega}\frac{1}{|\hbar\omega|^3}
\int\limits_{-\infty}^{+\infty}\!\! {\rm
d}\alpha\;\frac{\alpha^2}{(\alpha^2+1)^3}.
\end{align}
Here the integral is equal to $\pi/8$, and for the sum over the
Matsubara frequencies we obtain $\sum_{\omega}1/|\hbar\omega|^3
=7\zeta(3)/(4\pi^3T^3_c)\big(1+{\cal O}(\tau)\big)$. This gives
Eq.~(\ref{eq:Ia_4B_add}).

%%%%%%%%%%%%%%%%%%%%%%%%%%%%%%%%%%%%%%%%%%%%%%%%%%%%%%%%%%%%%%%%%%%%
\subsection{Terms related to $K_b({\bf r},\{y\}_3)$}
\label{subsec:C2}
%%%%%%%%%%%%%%%%%%%%%%%%%%%%%%%%%%%%%%%%%%%%%%%%%%%%%%%%%%%%%%%%%%%%

The second generation of the terms appearing in the r.h.s. of the
field-modified Eq.~(\ref{eq:gap_perturbation}) comes from
Eq.~(\ref{eq:int_Kb}), which can now be rewritten in terms of the
auxiliary ``two-point" order parameter, i.e., given by
Eq.~(\ref{eq:order_parameter_modified}), as
\begin{multline}
I_b=\lim\limits_{{\bf r}'\to {\bf
r}}\frac{1}{g}\int\!\!\prod\limits_{i=1}^3{\rm d}^3z_i\;
Q_b({\bf r},{\bf z}_1,{\bf z}_2,{\bf z}_3)\\
\times \Delta({\bf r}+{\bf z}_1,{\bf r}') \Delta^{\ast}({\bf r}+{\bf
z}_2,{\bf r}') \Delta({\bf r}+{\bf z}_3,{\bf r}'),
\label{eq:int_Kb_B}
\end{multline}
where
\begin{align}
Q_b({\bf r},{\bf z}_1,{\bf z}_2,{\bf z}_3) = K_{b,B=0}({\bf
z}_1,{\bf z}_2,{\bf z}_3)\;e^{\Sigma}
\end{align}
and $\Sigma=\Sigma({\bf r},{\bf z}_1,{\bf z}_2,{\bf z}_3)$ is given
by
\begin{align}
&\Sigma({\bf r},{\bf z}_1,{\bf z}_2,{\bf z}_3)=\nonumber\\
&=\Phi({\bf r},{\bf r}+{\bf z}_2)+\Phi({\bf r}+{\bf z}_2,{\bf
r}+{\bf z}_1)+\Phi({\bf r}+{\bf z}_1,{\bf r})\nonumber\\
&+\Phi({\bf r},{\bf r}+{\bf z}_2)+\Phi({\bf r}+{\bf z}_2,{\bf
r}+{\bf z}_3)+\Phi({\bf r}+{\bf z}_3,{\bf r}),\label{eq:SigPhi}
\end{align}
with $\Phi({\bf r},{\bf r}')$ given by Eq.~(\ref{eq:Phi}). Equation~(\ref{eq:SigPhi}) can be rewritten in the form
\begin{align}
\Sigma=\iint\limits_{S_1}{\bf B}({\bf r})\cdot{\rm d}{\bf S}_1 +
\iint\limits_{S_2}{\bf B}({\bf r})\cdot{\rm d}{\bf S}_2 + {\cal
O}(\tau^{3/2}), \label{eq:SigPhi_exp}
\end{align}
where the boundary of the surface $S_1$ consists of the three line
segments that connect the points ${\bf r}$, ${\bf r}+{\bf z}_1$, and
${\bf r}+ {\bf z}_2$, i.e., ${\bf r} \to {\bf r}+{\bf z}_1 \to {\bf
r}+{\bf z}_2 \to {\bf r}$; and $S_2$ is bounded by the three line
segments between the points ${\bf r}$, ${\bf r}+{\bf z}_3$, and
${\bf r}+{\bf z}_2$, i.e., ${\bf r} \to {\bf r}+{\bf z}_3 \to {\bf
r}+{\bf z}_2 \to {\bf r}$. Linear boundaries make it possible to
analytically calculate $\Sigma({\bf r},{\bf z}_1,{\bf z}_2,{\bf
z}_3)$, i.e.,
\begin{align}
\Sigma=\tau \Big(\bar{\bf B}(\bar{\bf r})\cdot \big[({\bf
z}_1\times{\bf z}_2) +({\bf z}_3\times{\bf z}_2)\big]\Big)+ {\cal
O}(\tau^{3/2}). \label{eq:SigPhi_exp_an}
\end{align}

Now, we have everything at our disposal to collect the relevant
terms up to ${\cal O}(\tau^{5/2})$ in
Eq.~(\ref{eq:gap_equation_series}). Based on the consideration in
Appendix \ref{sec:appendix_A}, one obtains
\begin{align}
&I^{(1)}_b=-b_1\tau^{3/2}\lim\limits_{{\bf r}'\to{\bf
r}}|\bar{\Delta}(\bar{\bf r},\bar{\bf r}')|^2\bar{\Delta}(\bar{\bf
r},\bar{\bf r}'), \label{eq:Ib_1a_A}\\
&I^{(2)}_b = - b_2\tau^{5/2}\lim\limits_{{\bf r}'\to{\bf r}} \Big[
2\bar{\Delta}(\bar{\bf r},\bar{\bf r}')\;|\bar{\bf \nabla}_{\bf r}
\bar{\Delta} (\bar{\bf r},\bar{\bf r}')|^2 \nonumber\\
&\quad+3\bar{\Delta}^\ast(\bar{\bf r},\bar{\bf r}')\big(\bar{\bf
\nabla}_{\bf r}\bar{\Delta}(\bar{\bf r},\bar{\bf r}') \big)^2
+\bar{\Delta}^2(\bar{\bf r},\bar{\bf r}')\; \nonumber\\
&\quad\times \bar{\bf \nabla}_{\bf r}^2\bar{\Delta}^{\ast}(\bar{\bf
r}, \bar{\bf r}') +4|\bar{\Delta}(\bar{\bf r},\bar{\bf r}')|^2
\bar{\bf \nabla}_{\bf r}^2\bar{\Delta}(\bar{\bf r},\bar{\bf
r}')\Big], \label{eq:Ib_1a_B}
\end{align}
which results in Eqs.~(\ref{eq:nabla_shiftedB}) (here $\bar{\nabla}$
can safely be replaced by $\bar{\bf D}$ in the relevant
expressions).

It may seem that there is one more term of the order $\tau^{5/2}$ coming
from the kernel $K_b$ in the presence of a nonzero magnetic field,
i.e.,
\begin{align}
&I^{(3)}_b=\frac{\tau^{5/2}}{g}|\bar{\Delta}|^2\bar{\Delta}\int\!\!
\prod\limits_{i=1}^3{\rm d}^3z_i\; K_{b,B=0}({\bf z}_1,{\bf
z}_2,{\bf z}_3)\nonumber\\
&\quad\quad\quad\quad\quad\times \Big(\bar{\bf B}(\bar{\bf r})\cdot
\big[({\bf z}_1\times{\bf z}_2) +({\bf z}_3\times{\bf
z}_2)\big]\Big).
\end{align}
However, we immediately find that $I^{(3)}_b=0$ due to
$K_{b,B=0}({\bf z}_1,{\bf z}_2,{\bf z}_3)=K_{b,B=0}(-{\bf z}_1,-{\bf
z}_2,-{\bf z}_3)$.

Concluding Appendix \ref{sec:appendix_C}, we note that the only term
appearing in Eq.~(\ref{eq:gap_equation_series}) from the integral
with the kernel $K_c$ is proportional to $|\bar{\Delta}|^4
\bar{\Delta}$ and, so, does not change its form in the presence of a
nonzero magnetic field (here corrections from ${\bf B}\not=0$ can
appear only in higher orders in $\tau$).
%%%%%%%%%%%%%%%%%%%%%%%%%%%%%%%%%%%%%%%%%%%%%%%%%%%%%%%%%%%%%%%%%%%%%

%%%%%%%%%%%%%%%%%%%%%%%%%%%%%%%%%%%%%%%%%%%%%%%%%%%%%%%%%%%%%%%%%%%%%%%%%%%%%%%%%%
%%%% References
%%%%%%%%%%%%%%%%%%%%%%%%%%%%%%%%%%%%%%%%%%%%%%%%%%%%%%%%%%%%%%%%%%%%%%%%%%%%%%%%%%


\begin{thebibliography}{99}
\bibitem{note1} In the "local superconductor" approach, the requirement of a small gap is relaxed and an expansion in the gradients of the order parameter is made.
\end{thebibliography}
\end{document}